\documentclass[twocolumngrid,prd,superscriptsize,reprint]{revtex4-1}
\usepackage[utf8]{inputenc} 
\usepackage{graphicx,xcolor,overpic,mathtools}
\usepackage{amsthm,amsmath,amssymb}
\usepackage[colorlinks]{hyperref}
\usepackage{textcomp}
\definecolor{coolblack}{rgb}{0.0, 0.18, 0.39}
\hypersetup{
    colorlinks = false,
   }
\PassOptionsToPackage{normalem}{ulem}
\usepackage{ulem} 
\usepackage{sidenotes}
\usepackage{bm}
\usepackage{url}
\usepackage{todonotes}
\usepackage{physics}
\usepackage[makeroom]{cancel}
\usepackage{cleveref}
\usepackage{tensor}
\usepackage{mathrsfs}
\usepackage{slashed}
\usepackage{caption}
\usepackage{subcaption}
\usepackage[mode=buildnew]{standalone}

\newcommand{\comment}[1]{}

\usepackage{pgfplots}

\usepackage{diagbox}
\usepackage{stackengine,amssymb,graphicx}

\NewDocumentCommand{\evat}{sO{\bigg}mm}{%
  \IfBooleanTF{#1}
   {\mleft. #3 \mright|_{#4}}
   {#3#2|_{#4}}%
}
\usepackage{enumerate}
\definecolor{azure}{rgb}{0.0, 0.5, 1.0}

\newcommand{\jc}[1]{{\color{green} #1}}
\begin{document}

\title[]{Universal Relations for Rotating Scalar and Vector Boson Stars}

\author{Christoph Adam}
\author{Jorge Castelo Mourelle}

\affiliation{%
Departamento de F\'isica de Part\'iculas, Universidad de Santiago de Compostela and Instituto
Galego de F\'isica de Altas Enerxias (IGFAE) E-15782 Santiago de Compostela, Spain
}%
\author{Etevaldo dos Santos Costa Filho}
\author{Carlos A. R. Herdeiro}
\affiliation{
Departamento de Matem\'atica da Universidade de Aveiro and Centre for Research and
Development in Mathematics and Applications (CIDMA),  Campus de Santiago, 3810-193
Aveiro, Portugal
}%

\author{Andrzej Wereszczynski}
\affiliation{
Institute of Physics, Jagiellonian University, Lojasiewicza 11, Krak\'ow, Poland
}%

\date[ Date: ]{\today}
\begin{abstract}

Bosonic stars represent a hypothetical exotic type of compact stellar objects that could be observed from the gravitational signal of coalescing binaries in current and future gravitational wave detectors. There are two main families of bosonic stars, which depend on the nature that governs the particles that build them: Einstein-Klein-Gordon and Proca Stars.
We study the multipolar structure for both families of rotating objects, using realistic potentials with the aim of finding possible universal relations and, thus, a method that allows us to distinguish between these and other compact objects in the gravitational wave paradigm.  We also show how 
certain relevant 
observables can be obtained for these hypothetical but well-motivated astrophysical objects.

\end{abstract}

\maketitle

\begin{quote}
 
\end{quote}

\tableofcontents

\section{Introduction}
Bosonic stars are localized solutions of a boson field theory coupled to gravity. In their simplest guise, these hypothetical astrophysical objects emerge in standard field theoretical models where the bosonic matter field is minimally coupled to gravity and described by a massive, free or self-interacting complex scalar (BS) \cite{PhysRev.172.1331,PhysRev.187.1767,PhysRev.148.1269,Schunck:2003kk} or a massive complex vector (Proca stars-PS) \cite{Brito:2015pxa,Herdeiro:2023wqf}. Recently, bosonic stars in a model with a complex vector coupled to a real scalar (Proca-Higgs stars-PHS \cite{Herdeiro:2023lze,Brito:2024biy}) which provides effective and consistent self-interactions of the Proca field were analysed. Other multi-field bosonic star models, such as multi-state boson stars \cite{Urena-Lopez:2010zva,Sanchis-Gual:2021edp} or $\ell$-Bosonic stars \cite{Alcubierre:2018ahf,Lazarte:2024jyr}, have also been investigated in recent years. 

The concept of bosonic stars originated from the proposals of geons and spherical BS \cite{PhysRev.97.511,PhysRev.172.1331,PhysRev.187.1767,PhysRev.148.1269}. Since then, the properties and phenomenology of these compact objects have been investigated in detail, and their formation mechanisms as well as their stability have been explored. Extensive reviews can be found, e.g., in  \cite{Liebling:2012fv,Lai:2004fw,Schunck:2003kk}. Properties of these kinds of stars strongly depend on the particular type of matter field and on a particular choice of the Lagrangian, i.e., the potential that encodes various types of self-interactions of the matter field. This allows for the modeling of various astrophysical systems, including Neutron Star (NS)-like objects, dark matter galaxy haloes, Black Hole mimickers, and intermediate-mass astrophysical objects \cite{Schunck:1998nq,PhysRevD.80.084023,Herdeiro:2021lwl,Rosa:2022tfv,Rosa:2023qcv,Sengo:2024pwk,Adam:2010rrj}. The scientific interest in the field has further increased due to the possible existence of dark-matter ultralight scalar bosons \cite{Freitas:2021cfi,2024arXiv240104735K} or extensions of the Standard Model, such as the axion \cite{PhysRevLett.40.223,PhysRevLett.40.279,Guo:2023hyp,Sakharov:2021dim}. 

From an astrophysical perspective, static, spherically symmetric BS are just the starting point. Realistic BS should be rather generically rotating objects. Rotation leads to axisymmetric spinning BS which have been found both for scalars \cite{Schunck:1996he,Yoshida:1997qf}, vectors \cite{Brito:2015pxa,Herdeiro:2016tmi,Herdeiro:2019mbz}, and PHS \cite{Herdeiro:2024pmv}. Such BS have been studied from a phenomenological point of view, e.g. \cite{Vincent:2015xta,Delgado:2021jxd,Delgado:2022pwo,Sengo:2024pwk,Sukhov:2024bwo,Vaglio:2023zpm}, and their stability as well as dynamical properties have been explored \cite{Sanchis-Gual:2019ljs,Sanchis-Gual:2021phr,Siemonsen:2020hcg}. 

In fact, dynamics of BS attracts more and more attention. Since the first event was reported by the LIGO-VIRGO collaboration \cite{LIGOScientific:2016aoc}, gravitational wave (GW) astronomy has become an increasingly powerful tool for studying the Universe. Advanced LIGO, Virgo, and KAGRA have now reported numerous events \cite{LIGOScientific:2021hvc,KAGRA:2018plz}, including binary black hole and neutron star mergers \cite{LIGOScientific:2018cki}, as well as mergers that suggest non standard interpretations. One such event observed in 2020 by advanced LIGO-Virgo could potentially be explained as a head-on collision of two Proca stars \cite{Bustillo:2020syj}. Even other events can be described under this same phenomenology, as shown in \cite{Luna:2024kof}. The dynamics of two boson stars  is an ongoing research program in full numerical relativity, see e.g. \cite{Palenzuela:2007dm,Sanchis-Gual:2018oui,Palenzuela:2006wp,Palenzuela:2007dm,Sanchis-Gual:2018oui,Bezares:2022obu}, albeit offering significant challenges~\cite{Sanchis-Gual:2022mkk,Siemonsen:2023age}. The construction of large libraries of waveforms for data comparison is a costly task and may be tackled by novel strategies, such as neural networks~\cite{Freitas:2022xvg,Luna:2024kof}.  Undoubtedly, the possibility of boson stars as compact astrophysical sources distinct from black holes and neutron stars is a growing research area in the GW community \cite{Cardoso:2019rvt,LIGOScientific:2021sio,Maggio:2021ans,CalderonBustillo:2020fyi}. 

Both black holes as well as neutron stars give rise to the so-called Universal Relations. In the case of black holes it is just the no-hair theorem \cite{misner1973gravitation,Robinson:1975bv,Israel:1967wq,Hawking:1971tu,hawking1972black,Carter:1971zc}, which states that in electrovacuum General Relativity these compact objects are completely characterized by their mass, angular momentum and charge. In the case of neutron stars, whose properties strongly depend on details of the particular Equation of State of the nuclear matter, universality manifests on the level of relations between some quantities. The most famous are the $I$-Love-$Q$ relations, proposed by Yagi and Yunes in \cite{Yagi:2013awa}, describing the universal, Equation of State independent, relationships between the moment of inertia $I$, tidal deformability (Love number) \cite{Hinderer:2007mb,Postnikov:2010yn}, and quadrupolar moment $Q$. Following this pioneering work many other universal and quasi-universal relations for NS have been identified - see e.g., extension to high spins and magnetic neutron stars \cite{Haskell:2013vha}, as well as extension  to modified gravity theories \cite{Sham:2013cya,Chakravarti:2019aup,Doneva:2017jop}. Other quasi-universal relations, including those involving higher multipoles and Love numbers \cite{Yagi:2013sva,Godzieba:2021vnz}, compactness, gravitational binding energies \cite{UnivRelsbinding}, and oscillation frequencies of (quasi)normal modes \cite{Torres-Forne:2019zwz}, have been studied. 

The universal relations of NS are believed to be related to the no-hair theorem of black holes and therefore are often referred to as effective no-hair relations. Although this statement still requires a detailed mathematical proof, it is motivated by the observation that universal relations usually concern quantities obtained in multipole expansions. A Kerr BH exterior gravitational field can be  reconstructed as an infinite series of multipoles, depending only on the mass-monopole and the current-dipole \cite{Geroch:1970cd,Hansen:1974zz} (for beyond vacuum cases, see \cite{Simon,Pappas:2014gca,Fodor:2020fnq,Filho:2021exr,10.21468/SciPostPhys.15.4.154}). The importance of these multipoles lies not only in their link with the gravitational field created by an object but also  in their direct relation with astrophysical observables \cite{Ryan2,Ryan3,Pappas:2012nt}. For NS (and Quark Stars) the BH no-hair theorems do not apply, as they are non-vacuum sources, but universal and quasi-universal relations do exist and connect various multipoles. Their interpretation as effective no-hair theorems for fermionic compact objects was treated in several works \cite{Yagi:2014bxa,Doneva:2017jop,Stein:2013ofa,Yagi:2013awa,Yagi:2016bkt}. 

Importantly, the universality is not only an interesting mathematical property but it also provides a useful tool for studying quantities which are difficult to extract in observational data.  These relations also aid in breaking the degeneracy between the NSs' spin parameter and the quadrupolar moment in binary systems \cite{Yagi:2013awa,Yagi:2016bkt}. 

A natural question arises whether also for other compact objects, especially BS, some universal relations exist. This is, in fact, the case. It has recently been  demonstrated that, despite the variety of self-interaction potentials, the complex scalar field boson stars follow their own universal relations involving the first two multipoles \cite{Adam:2022nlq,Vaglio:2022flq}. Interestingly, BS and NS give rise to functionally different universal relations involving the same quantities. 

Recently, integral and universal relations have been found in solitonic theories without coupling to gravity, linking observables related to the quadrupole and inertia moments, suggesting that this phenomenon might be more general than expected \cite{Adam:2024cem}.

\vspace*{0.2cm}

In this paper we further investigate the problem of effective-no-hair relations for bosonic stars. In particular we ask the question whether different target spaces of the matter fields lead to different quasi-universal relations. We investigate this problem using scalar BS, Proca stars and Proca-Higgs stars. In particular, we are interested in relations among the multipole moments up to the octupolar order. 

Contrary to previous works, we will test such relations also for rapidly rotating compact stars.

The structure of the paper is the following. In  \cref{setup} we introduce
the theoretical set-up for the models under study. We present the numerical scheme in  \cref{numerical}.  In \cref{multi} we show how to obtain the multipolar expansion for our stationary and axisymmetric space-time systems and other observables of interest like the moment of inertia $I$. In \cref{results} we present results concerning the universal behaviors and comparisons with other compact objects. Section \ref{conclusions} is devoted to conclusions, whereas some relevant equations and the numerical values of some parameters and fitting constants are shown in \cref{appendix}.


\section{Theoretical set-up}\label{setup}
\subsection{Scalar Boson Stars}
The scalar boson stars considered in this paper are described by the Einstein-Klein-Gordon (EKG) action, where a  massive complex scalar field $\Phi$ is minimally coupled to Einstein gravity \cite{Liebling:2012fv},
\begin{equation}
    \mathcal{
    S}_{\Phi}=\int \left(\frac{1}{16\pi G}R+\mathcal{L}_{\Phi}\right)\sqrt{-g}d^4x.
    \label{action}
\end{equation}
Here, $g$ is the metric determinant, $R$ the Ricci scalar, and the Lagrangian  governing the field dynamics reads
\begin{equation}
 \mathcal{L}_{\Phi}=-\frac{1}{2}\left[g^{\alpha\beta}\nabla_{\alpha}\Phi^*\nabla_{\beta}\Phi+V\left(|\Phi|^2\right)\right] ,
    \label{lagrangian}
\end{equation}
see, e.g., \cite{Adam:2022nlq,Adam:2023qxj} for details.
Respecting global $U(1)$ invariance of the model, the potential $V\left(|\Phi|^2\right)$ depends only on the absolute value of the scalar field. Each model has a different self-interacting term, allowing us to get various types of BS. All potentials we consider contain the quadratic mass term $\mu^2|\Phi|^2$, plus some additional interactions. The scalar potential for the BS plays an analogous role to the EOS in the NS case. In this work, we have used the same models as in \cite{Adam:2023qxj}, and detailed potentials with their corresponding coupling constants are shown in \cref{appendix}. 

By varying the action (\ref{action}) the EKG equations arise,
\begin{equation}
\begin{split}
    &R_{\alpha\beta}-\frac{1}{2}Rg_{\alpha\beta}=8\pi T^s_{\alpha\beta}, \\
    &
    g^{\alpha\beta}\nabla_\alpha\nabla_{\beta}\Phi=\frac{dV}{d|\Phi|^2}\Phi,
\label{kg}
\end{split}
\end{equation}
where $R_{\alpha\beta}$ is the Ricci tensor and $T^s_{\alpha\beta}$ is the canonical Stress-Energy tensor of the scalar field,
\begin{equation}
    T_{\alpha\beta}^s=\nabla_{(\alpha}\Phi^*\nabla_{\beta)}\Phi-g_{\alpha\beta}\left[\frac{1}{2}g^{\mu\nu}\nabla_{(\mu}\Phi^*\nabla_{\nu)}\Phi+V\left(|\Phi|^2\right)\right]].
    \label{stress}
\end{equation}
For the above Stress-Energy tensor to satisfy stationarity and axial symmetry, the scalar field ansatz takes the form
\begin{equation}
     \Phi(t,r,\theta,\psi)=\phi(r,\theta)e^{-i(w t+n\psi)}.
     \label{scalar}
 \end{equation}
Here $w \in \mathbb{R}$ is the angular frequency of the field, and $n \in \mathbb{Z}$ (also called $m$ or  $s$ in the literature \cite{Vaglio:2022flq,Ryan:1996nk}) is the \textit{azimutal harmonic index}, also called \textit{azimutal winding number}. This parameter enters the problem as an integer related to the star's angular momentum. Further, $\phi(r,\theta)$ is the profile of the star.
 Finally, we assume the following ansatz for the metric, describing the stationary 
and axisymmetric space-time\cite{Herdeiro:2015gia,PhysRevD.55.6081},
\begin{equation}
\begin{split}
    ds^2=&-e^{2\nu}dt^2+e^{2\beta}r^2\sin^2\theta\left(d\psi-\frac{W}{r}dt\right)^2\\
    &+e^{2\alpha}(dr^2+r^2d\theta^2),
    \end{split}
    \label{metric}
\end{equation}
where $\nu, \alpha, \beta$ and $W$ are functions dependent only on $r,\theta$. 
 
While the universal I-love-Q relation has been discovered in the context of NS, a number of related universal relations for the known scalar boson stars was shown to exist, e.g., in \cite{Adam:2023qxj}.

\subsection{Vectorial  Boson Stars}
In this section we follow \cite{Herdeiro:2016tmi,Brito:2015pxa}.
The Einstein-Proca model is described by the action
\begin{equation}
    \textit{S}_P=\int d^4x\sqrt{-g}\left(\frac{R}{16\pi G}-\frac{1}{4}\textit{F}_{\alpha\beta}\bar{\textit{F}}^{\alpha\beta}-\frac{1}{2}\mu^2\textit{A}_{\alpha}\bar{\textit{A}}^{\alpha}\right) ,
\end{equation}
where $F_{\alpha\beta}=\partial_{\alpha}A_{\beta}-\partial_{\beta}A_{\alpha}$, and the Einstein-Proca system is governed by the equations
\begin{equation}
\begin{split}
    &R_{\alpha\beta}-\frac{1}{2}Rg_{\alpha\beta}=8\pi T^p_{\alpha\beta}, \\
    &
    \nabla_{\alpha}\textit{F}^{\alpha\beta}=\mu^2\textit{A}^{\beta},
\label{pr}
\end{split}
\end{equation}
implying the Lorenz condition $\nabla_{\alpha}\textit{A}^{\alpha}=0$.
The explicit equations of motion with the Christoffel symbols are
\begin{equation}
\textit{F}^{\alpha\beta}_{,\alpha}+\Gamma^{\alpha}_{\mu\alpha}\textit{F}^{\mu\beta}+\Gamma^{\beta}_{\mu\alpha}\textit{F}^{\alpha\mu}-\mu^2A^{\beta}=0,
\end{equation}
and the Lorenz condition reads
\begin{equation}
    \mathcal{L}=A^{\alpha}_{,\alpha}+\Gamma^{\alpha}_{\mu\alpha}A^{\mu}=0.
    \label{lorenz}
\end{equation}

As we look for spinning solutions, we use the same metric \cref{metric} describing the axisymmetric spacetime as for the scalar case. The Stress-Energy tensor takes the form,
\begin{equation}
\begin{split}
    T_{\alpha\beta}=&-F_{\sigma(\alpha}\bar{F}_{\beta)}^{\sigma}-\frac{1}{4}g_{\alpha\beta}F_{\mu\nu}\bar{F}^{\mu\nu}+\\
    &
    \mu^2\left[A_{(\alpha}\bar{A}_{\beta)}-\frac{1}{2}g_{\alpha\beta}A_{\sigma}\bar{A}^{\sigma}\right],
\end{split}
\end{equation}

The field ansatz that closes the problem is
\begin{equation}\label{ansatzProca}
    \textit{A}=e^{-i(n\psi+w t)}\left(i V dt+\frac{H_1}{r}dr+H_2 d\theta+iH_3\sin\theta d\psi \right) ,
\end{equation}
where $A= A_\mu dx^\mu$ is the one-form corresponding to the co-vector $A_\mu$. Further, $V,H_1,H_2$ and $H_3$ are functions depending on $r$ and $\theta$.
In the current paper, we only consider Proca Stars with harmonic index $n=1,2$.

Recent studies reveal that introducing self-interactions in Proca fields, while theoretically grounded in various phenomenological contexts, leads to critical challenges in field consistency  \cite{Coates:2022nif,PhysRevLett.129.151102,Mou:2022hqb,PhysRevLett.129.151103,Barausse:2022rvg,Brito:2024biy}. Predominant issues include hyperbolicity and emerging instabilities. This contrasts with the scalar field case, where self-interactions do not typically induce such fundamental issues. Due to these inherent problems, self-interacting potentials in Proca fields are outside our research scope.

This factor complicates the analysis by precluding a direct comparison within PS when attempting to elucidate the purported existence of universal relations.

\subsection{Proca-Higgs Stars}

We also consider a scalar-vector model in which the Einstein gravity is minimally coupled to a real scalar field $\Psi$, and a complex vectorial field  $A_{\alpha}$. 
This model was initially explained and analyzed in a static context in \cite{Herdeiro:2023lze} and within the axisymmetric framework in \cite{Herdeiro:2024pmv}. It can also be seen as an ultraviolet completion of a self-interacting Proca model. In this system, the mass of the vector field is provided by the scalar-vector coupling and, more concretely, by the nontrivial vacuum expectation value (vev) of the scalar field. The action describing this model is
\begin{equation}
\begin{split}
    \textit{S}_{PH}=&
    \int d^4x\sqrt{-g}\left(\frac{R}{16\pi G}-\frac{1}{4}\textit{F}_{\alpha\beta}\bar{\textit{F}}^{\alpha\beta}-\frac{1}{2}\Psi ^2\textit{A}_{\alpha}\bar{\textit{A}}^{\alpha}\right.\\&
    \left.-\frac{1}{2}\partial_{\alpha}\Psi\partial^{\alpha}\Psi-\mathcal{U}(\Psi)\right),
    \end{split}
\end{equation}
where $\mathcal{U}$ is the double vacuum $\phi^4$ potential,
\begin{equation}
    \mathcal{U}(\Psi)=\frac{\Lambda}{4}(\Psi^2-v^2)^2, 
\end{equation}
$v$ is the vev, and $\Lambda$ is a positive constant. Taking all the above into account, we can describe the system with the following set of equations:
\begin{equation}
    \begin{split}
        &\nabla_{\alpha}F^{\alpha\beta}=\Psi^2A^{\beta},\\&
    g^{\alpha\beta}\nabla_\alpha\nabla_{\beta}\Psi=\frac{d\mathcal{U}(\Psi)}{d\Psi}+\Psi A_{\alpha}A^{\alpha},\\&
     R_{\alpha\beta}-\frac{1}{2}Rg_{\alpha\beta}=8\pi( T^1_{\alpha\beta}+T^2_{\alpha\beta}),
    \end{split}
\end{equation}
and the stress-energy momentum expressions are,
\begin{equation}
    \begin{split}
      & T^1_{\alpha\beta}= -F_{\sigma(\alpha}\bar{F}_{\beta)}^{\sigma}-\frac{1}{4}g_{\alpha\beta}F_{\mu\nu}\bar{F}^{\mu\nu}+\\
    &
    \Psi^2\left[A_{(\alpha}\bar{A}_{\beta)}-\frac{1}{2}g_{\alpha\beta}A_{\sigma}\bar{A}^{\sigma}\right],\\&
       T^2_{\alpha\beta}=\partial_{\alpha}\Psi\partial_{\beta}\Psi-g_{\alpha\beta}\left[\frac{1}{2}(\partial\Psi)^2+\mathcal{U}(\Psi)\right].
    \end{split}
\end{equation}

The Lorenz-like condition is again a consequence of the equations of motion but now takes the form: $\nabla_\alpha\left(\phi^2A^\alpha\right)=0$. The usual Einstein-Proca model is a limit when $\Lambda\rightarrow \infty$, forcing the Higgs field to be fixed at the vev.
In what follows, we are going to treat the Einstein-Proca-Higgs models as different families of Proca Stars. It is critical to note that, although the system adheres to the characteristics of the PS asymptotically, the presence of an additional scalar field renders the object a hybrid scalar-vector star. Even so, formally, we treat these objects as a type of Proca Stars.

The field ansatz for the vector part is the same as in the pure Proca case (see Eq. \eqref{ansatzProca}). As for the scalar part, we take $\Psi=\Psi(r,\theta)$.
\section{Numerical implementation }\label{numerical}

Even though, at the level of the code, there are many subtleties, the philosophy and methods used for performing the numerical integration of our different systems are the same. For Boson and Proca stars, we rescale the radial distance and angular frequency by the mass $\mu$ of the boson field, $
r\rightarrow r\mu, \hspace{0.2cm}w\rightarrow w/\mu$ which eliminates the the explicit $\mu$ dependence from the field equations but changes the coupling constant definitions for the different potentials in the scalar case. For the Proca-Higgs stars, in turn, we use the vev for the rescaling, i.e.,  $r\rightarrow \dfrac{r}{v}\,,  \omega\rightarrow v \omega \,,\phi\rightarrow v\phi\, ,   \mathcal{A}_{\alpha}\rightarrow v \mathcal{A}_{\alpha}$, and we define the new parameter $\alpha^2\equiv 4\pi G v^2$.

We have three problems, each one with a different set of coupled, non-linear, partial differential equations. There are five of them for the metric functions and the scalar field in the scalar case, eight equations for the four metric plus four fields in the common vectorial case, and nine for the Proca-Higgs models. We also take into account the constraints, $E^r_{\theta}=0, E^r_r-E^{\theta}_{\theta}=0$, where $E^{\mu}_{\nu}=R^{\mu}_{\nu}-\frac{1}{2}Rg^{\mu}_{\nu}-2 T^{\mu}_{\nu}$. We use the FIDISOL/CADSOL package \cite{fidisol,schonauer1989efficient,schonauer2001we}, a Newton-Raphson-based code with an arbitrary grid and
consistency order. It also provides an error estimate for each unknown function. 

We compactify the radial coordinate by the following definition $x\equiv r/(c+r)$ moving from $r\in [0,\infty)$ to a finite segment $x \in [0,1]$. For the BS, the discretization of the equations was done on a $(401\times 40)$, $(x,\theta)$ grid, where $0\leq x\leq 1$ and $0\leq \theta\leq \pi/2$. The considerable size of the grid allows us to fix $c=1$. The Proca case is more expensive in terms of computing time, so the size of the grid was reduced, but as it is important to have enough points in the far field region, we adjust the $c$ value depending on the frequency regions where we are obtaining solutions.

\subsection{Scalar BS boundary conditions}
We impose boundary conditions on the field profile and the metric functions. Asymptotic flatness leads to,
\begin{equation}
    \lim_{r\rightarrow\infty} \alpha =\lim_{r\rightarrow\infty} \beta =\lim_{r\rightarrow\infty} \nu =\lim_{r\rightarrow\infty} W =\lim_{r\rightarrow\infty} \phi = 0.
\end{equation}
Reflection on the rotation axis and  axial symmetry imply that at $\theta=0$ and $\theta=\pi$,
\begin{eqnarray}
    \partial_{\theta}\alpha=\partial_{\theta}\beta=\partial_{\theta}\nu=\partial_{\theta}W=\partial_{\theta}\phi=0. 
\end{eqnarray}
Since the solutions have to be symmetric with respect to a reflection along the equatorial plane, this condition is also obeyed on the equatorial plane, $\theta=\pi/2$.
Eventually, regularity at the origin requires $\partial_r \alpha=\partial_r \beta=\partial_r \nu =W= \phi=0$ when $r \to 0$, and regularity in the symmetry axis further imposes $\left.  \alpha=\beta \right|_{\theta=0, \pi}$ \cite{Herdeiro:2015gia}.
Further details about the solver are explained in  \cite{Delgado:2022pwo,Adam:2022nlq}.

\subsection{Vectorial BS boundary conditions}

In the vectorial case, we have to impose, at the origin, 
\begin{equation}
\begin{split}
&\partial_r\alpha|_{r=0}=\partial_r\beta|_{r=0}=\partial_r\nu|_{r=0}= \\&
    W|_{r=0}=H_i|_{r=0}=V|_{r=0}=0, 
   \end{split}
\end{equation}
and at infinity 
 \begin{equation}
 \begin{split}
   &\lim_{r\rightarrow\infty} \alpha=\lim_{r\rightarrow\infty} \beta=\lim_{r\rightarrow\infty} \nu=\\
    &\lim_{r\rightarrow\infty}W=\lim_{r\rightarrow\infty}H_i=\lim_{r\rightarrow\infty}V=0.
   \end{split}
 \end{equation}
On the symmetry axis, the boundary conditions are
\begin{equation}
\begin{split}
&\partial_{\theta}\alpha|_{\theta=0,\pi}=\partial_{\theta}\beta|_{\theta=0,\pi}=\partial_{\theta}\nu|_{\theta=0,\pi}=\\
&\partial_{\theta}W|_{\theta=0,\pi}=\partial_{\theta}H_{2,3}|_{\theta=0,\pi}=\\&H_1|_{\theta=0,\pi}=V|_{\theta=0,\pi}=0. 
\end{split}
\end{equation}

In addition to the above boundary conditions, for the Proca-Higgs case, we have also to consider the scalar part.
\begin{equation}
     \partial_r\Psi|_{r=0}=\partial_{\theta}\Psi|_{\theta=0,\pi/2}=0\,,\qquad \lim_{r\rightarrow\infty} \Psi=1\,.
\end{equation}

\begin{figure*}[]
\subfloat{%
  \hspace*{0.9cm}\includegraphics[clip,width=1.0\columnwidth]{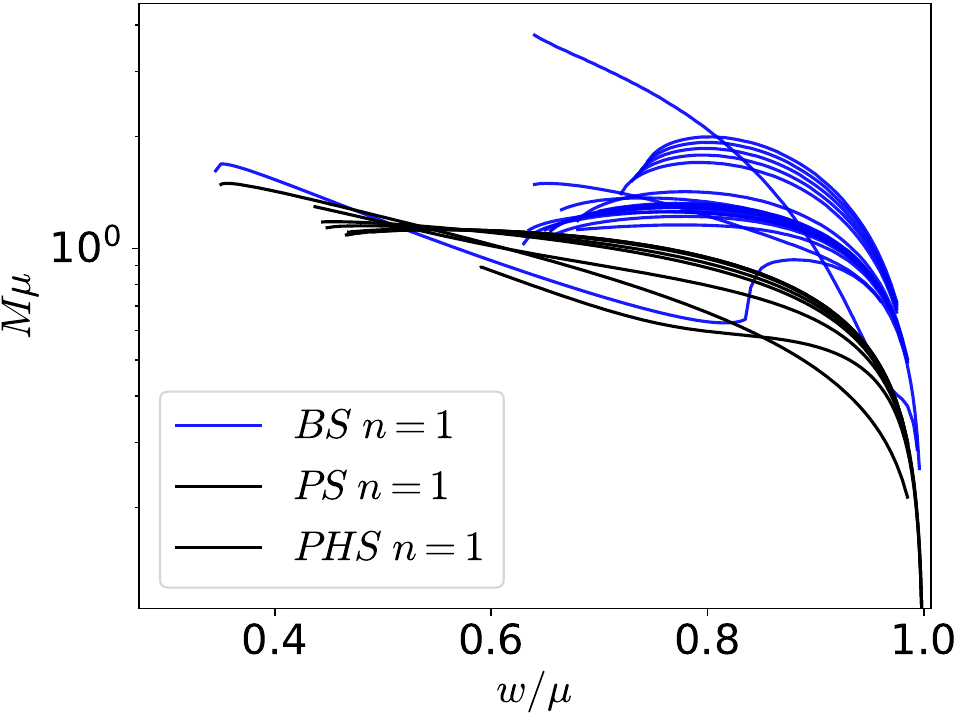}%
}

\subfloat{%
  \hspace*{0.9cm}\includegraphics[clip,width=1.0\columnwidth]{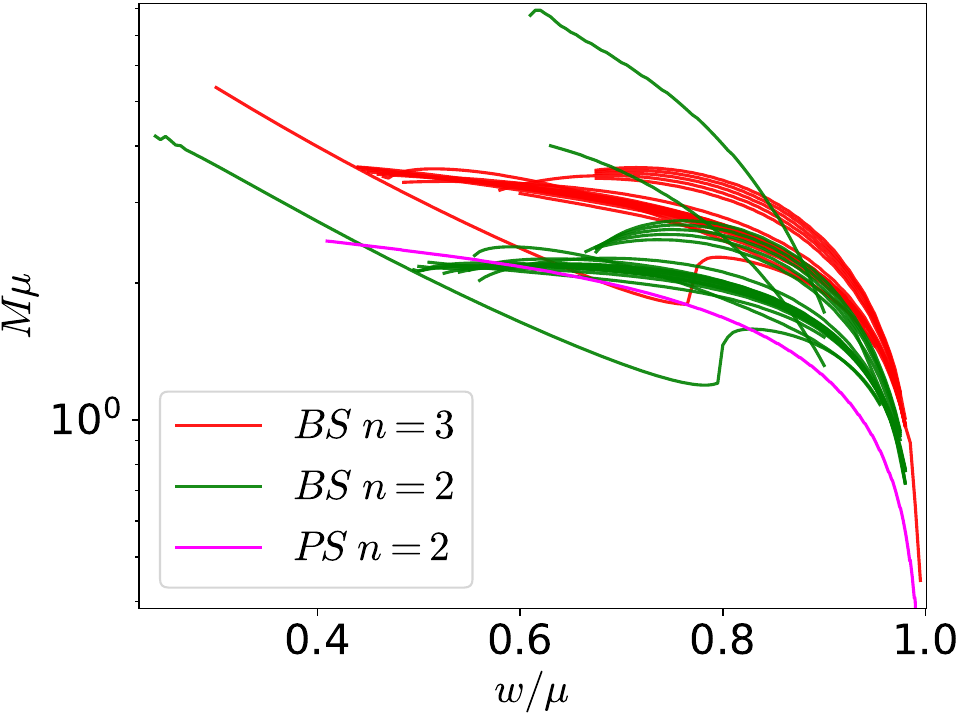}%
}
\caption{Mass frequency curves for the different models we use,  all displayed in \cref{appendix}. The upper plot shows the solutions for harmonic index  $n=1$ for the BS, PS, and PHS cases. The lower plot shows $n=2$ PS and BS and $n=3$ BS. The different potentials for obtaining those representative curves are shown in the appendix. As can be seen from the plot, we have some secondary branches and very different curves, making our set very wide-ranging. }
\label{mbS}
\end{figure*}

As can be read from \cref{mbS}, we have stars ranging from $\sim 0.2 M\mu$, to  $\sim 6 M\mu$ for some potentials. These more massive stars were already studied in detail in \cite{Ontanon:2021hbg}.

\section{Multipolar structure and global properties}\label{multi}

In this section, we study the space-time multipole expansion and the related physical properties relevant for the data analysis. Within the framework of General Relativity, there are two classes of multipoles, originating from the energy density and the current density, respectively, as discussed in \cite{RevModPhys.52.299,PhysRevD.52.821}. These multipoles are indispensable from both theoretical and astrophysical perspectives. The foundational principles of metric-multipole-expansions were established in \cite{Geroch:1970cd,Hansen:1974zz,fodor1989multipole}. Subsequently, this concept was applied within the context of neutron stars, as referenced in \cite{Pappas:2018csu,1976ApJ...204..200B}. Our approach is consistent with these precedents, however, it specifically addresses our scalar and vector boson scenario.

\subsection{Multipole moments}
In the same fashion as in \cite{Adam:2023qxj}, we obtain the multipoles following the method described in \cite{1976ApJ...204..200B,morse1954methods}. 
First we reparametrize our metric functions as $\omega=\frac{W}{r},\hspace{0.4cm}B=e^{\nu+\beta}$ and we expand our new metric functions using polynomial bases for spherical coordinates, that is, the  Legendre polynomials $P_l(\cos\theta)$ and the Gegenbauer polynomials $T_l^{\frac{1}{2}}(\cos\theta)$. The radial coefficients are also expanded in radial powers,
\begin{eqnarray}
        \nu&=&\sum_{l=0}^{\infty}\bar{\nu}_{2l}(r)P_{2l}(\cos\theta),\hspace{0.4cm}   \hspace{0.9cm} \bar{\nu}_{2l}(r)= \sum_{k=0}^{\infty}\frac{\nu_{2l,k}}{r^{2l+1+k}}, \nonumber  \\
         \omega&=&\sum_{l=0}^{\infty}\bar{\omega}_{2l-1}(r)\frac{dP_{2l-1}(\cos\theta)}{d\cos\theta},     \hspace{0.2cm} \bar{\omega}_{2l-1}(r)= \sum_{k=0}^{\infty}\frac{\omega_{2l-1,k}}{r^{2l+1+k}} \nonumber  \\
          B&=&1+\sum_{l=0}^{\infty}\bar{B}_{2l}(r)T_{2l}^{\frac{1}{2}}(\cos\theta),     \hspace{0.2cm}     \hspace{0.2cm} \bar{B}_{2l}(r)= \frac{B_{2l}}{r^{2l+2}},       
    \label{multipoles}
\end{eqnarray}

For more details we refer to \cite{Adam:2023qxj,PhysRevD.55.6081,Doneva:2017jop,Pappas:2012ns,Pappas:2012qg}. Once we know the functions $\nu$, $\omega$ and $B$, the coefficients are extracted straightforwardly. In the present study, we focus on the first few multipole moments, namely, the mass monopole $M_0$, spin $S_1$, mass quadrupole $M_2$, and spin octupole $S_3$. In terms of the metric coefficients, they are expressed as follows \cite{Pappas:2012ns,Pappas:2012qg}


\begin{equation}
\begin{split}
 &M_0=M = -\nu_{0,0},\\
 &S_1=J =\frac{\omega_{1,0}}{2},\\
 &M_2=Q=\frac{4}{3}B_{0}\nu_{0,0}+\frac{\nu_{0,0}^3}{3}-\nu_{2,0},\\
 &S_3=-\frac{6}{5}B_0\omega_{1,0}-\frac{3}{10}\nu^2_{0,0}\omega_{1,0}+\frac{3}{2}\omega_{3,0}.
    \end{split}
    \label{multipolcoeffs}
\end{equation}

Higher order multipole moments can be evaluated using algorithms as described in \cite{Geroch:1970cd,Hansen:1974zz,fodor1989multipole,Backdahl:2005be,Pappas:2012ns,Pappas:2012qg}.


\subsection{Moments of inertia and differential rotation}
As shown in \cite{Adam:2022nlq,Adam:2023qxj}, and also discussed in \cite{DiGiovanni:2020ror}, unlike regular perfect fluid stars, rotating BS require a full-rotating treatment\cite{Silveira:1995dh,Ferrell:1989kz}. We take advantage of the fact that there is a natural four-vector associated with the global $U(1)$ symmetry of the Lagrangian, i.e., the corresponding Noether current, that for scalar boson stars is,
\begin{equation}
    j^{\mu}_s=\frac{i}{2}\left[\Phi^*\partial^{\nu}\Phi-\Phi\partial^{\nu}\Phi^*\right],
\end{equation}
and for all vectorial stars,
\begin{equation*}
    j^{\mu}_p=\frac{i}{2}\left[\Bar{F}^{\mu\beta}A_{\beta}-F^{\mu\beta}\Bar{A}_{\beta}\right] .
\end{equation*}
These currents give rise to the conserved particle number,
\begin{equation}
  N_{(s,p)}=\int  j^0_{(s,p)} \sqrt{-g}d^3x  ,
\end{equation}
where the subindex $(s,p)$ labels the different kinds of stars.
Explicitly, the BS differential angular velocity is, 
\begin{equation}
\begin{split}
    \Omega_s=\frac{j^{\psi}_s}{j^{t}_s}=\frac{wg^{\psi t}-ng^{\psi\psi}}{wg^{tt}-ng^{t\psi}}=\frac{W}{r}+\frac{ne^{2(\nu-\beta)}}{r^2\left(w-\frac{nW}{r}\right)\sin^2\theta},
\label{diferentialfrequency}
\end{split}
\end{equation}
whereas the PS case, 
\begin{equation}
\begin{split}
    \Omega_p=\frac{j^{\psi}_p}{j^{t}_p},
\label{diferentialfrequencyP}
\end{split}
\end{equation}
leads to a much more complicated expression, which we show in  \cref{appendix}.
The above result, for the scalar case,  agrees with that obtained by Ryan in \cite{PhysRevD.55.6081} in the strong coupling approximation.
Having $\Omega_{(s,b)}$ as functions of $r$ and $\theta$, we compute the inertia tensor for our differentially rotating systems, 
\begin{equation}
    I_{(s,p)}=\int_0^{\pi}\int_0^{\infty}\frac{T^t_{\psi}(r,\theta)_{(s,p)}}{\Omega(r,\theta)_{(s,p)}}r^2\sin\theta e^{\nu+2\alpha+\beta}drd\theta \ .
\end{equation}

\section{Analysis}\label{results}

The reduced multipoles are standard definitions commonly used throughout the literature \cite{2013Sci...341..365Y,Yagi:2014bxa} and, as in \cite{Adam:2022nlq,Adam:2023qxj}, we work with these dimensionless quantities, which are completely general for any order
\begin{equation}
\begin{split}
    &m_{2n}\equiv (-1)^n\frac{M_{2n}}{\chi^{2n}M_0^{2n+1}}\\
    &s_{2n-1}\equiv (-1)^{n+1}\frac{S_{2n-1}}{\chi^{2n-1}M_0^{2n}}.
\end{split}
\end{equation}

In the above expression, $n$ may assume any integer value from $0$ to $\infty$, providing a consistent formula that removes the dimensions for each multipolar order. These equations are closely connected to the multipolar expansion method, which derives the expressions found in \cref{multipolcoeffs}. While the comprehensive calculation from the metric \cref{metric}, through the expansion of metric functions in \cref{multipoles}, is technically elaborate, the entire procedure is detailed in \cite{Pappas:2012ns,Pappas:2012qg} and meticulously reviewed in section $2.2 $ of \cite{Sukhov:2023rln}.

To work with these definitions, and in view of the absence of a surface for bosonic stars, we shall use $|M_0| \rightarrow |M_{99}|$ as the relevant mass parameter describing a star, with $M_{99}$ representing $99\%$ of the aggregate mass of the star. Then, the dimensionless spin parameter is defined as $\chi \equiv S_1/M_{99}^2$. Our \emph{reduced relevant physical quantities}, the moment of inertia, mass quadrupole, dimensionless spin and spin octupole, respectively, are thus

\begin{equation}
\begin{split}
    \bar{I}=&\frac{I}{M_{99}^3},\hspace{0.4cm}\bar{Q}=m_2=\frac{M_2}{M_{99}^3\chi^{2}},\hspace{0.1cm}\\
    &\chi=\frac{S_1}{M_{99}^2},
    \hspace{0.4cm}s_3=-\frac{S_3}{M_{99}^4\chi^4}.
    \label{reduced}
\end{split}
\end{equation}

The concept of compactness plays a pivotal role here, although it is not the central focus of most contemporary studies. Compactness, denoted as $C$ is defined herein as the ratio of the mass $M_{99}$ to the (circumferential) radius $R_{99}$ containing such mass, and given by the equation
\begin{equation}
C = \frac{M_{99}}{R_{99}}.
\end{equation}

Historically, the first universal behaviors associated with compact stars were identified in the 1990s, as evidenced in  \cite{Lattimer:1989zz}. This research established a correlation between the binding energy and the compactness of NSs. Subsequent studies in the same period, such as \cite{Andersson:1996pn,Andersson:1997rn}, explored relationships between the frequencies of 
$f-modes$, $w-modes$, and compactness, expanding our understanding to include the NS damping time for both modes and compactness.
Investigations into the link between compactness and the NS moments of inertia have been conducted, as documented in \cite{Lattimer:2000nx,Lattimer:2004nj,Bejger:2002ty}. Additionally, the connection between compactness and the quadrupole moments was established in \cite{Urbanec:2013fs}.

Recent research has focused on the relation of compactness with both the moment of inertia and the quadrupole moment of astrophysical objects. This line of inquiry led to the exploration of relationships between these two parameters. The field then expanded significantly, encompassing a diverse range of universal relations pertaining to neutron and quark stars. Key contributions in this area are \cite{Doneva:2017jop,Yagi:2016bkt,Yagi:2014bxa,Yagi:2013awa}.

In the context of BS, however, the exploration of universal behaviors and effective no-hair properties has been less extensive, primarily due to the more exotic nature of these stars. Previous research has touched upon this topic \cite{Adam:2023qxj}, leading to results in agreement with other notable studies in this realm \cite{Ryan:1996nk,Vaglio:2022flq,Grandclement:2014msa,Sukhov:2024bwo}. These works have contributed to a broader understanding of BS, albeit to a lesser extent compared to more conventional astrophysical objects.
  
Proca stars were previously examined in the context of compactness as outlined in \cite{Herdeiro:2016tmi}. It is important to note that these studies employed a distinct definition of compactness and were conducted within an alternative framework. But we will treat here PS as we did with BS in our previous works for the sake of performing the same kind of study, which is completely novel. 

For the physical quantities/multipole moments $\Bar{I},\chi,\Bar{Q},s_3$ and $C$, $n=1,2,3$ BS, $n=1$ PS and PHS, and $n=2$ PS data, we represent our simulations in 3D spaces where the different multipole moments will play the role of the dimensions. As was shown in \cite{Adam:2023qxj}, we can find a given surface in each triad of moments for BS having well-defined fitting surfaces, with a fitting error about  $\sim 7\%$ for the worst case and under $\sim 2\%$ in general. What we outline here is that the PS data, for each tetrad of multipoles, lie in different but comparable regions. The models of vector stars analyzed in this study occupy a shared region within the three-dimensional multipole space, and our findings indicate that the PS, together with the PHS models, conform to a region amenable to fitting, comparable to the behavior observed in the BS paradigm. Although our analysis is limited to only two families and does not encompass a broader range of models, it is noteworthy that our dataset allows us to find fitable surfaces with an accuracy of better than $11\%$, in general.

Putting our results within the well-studied context of NS, we observe that the various families of existing universal behaviors exhibit significant diversity in related observables/multipoles and in the precision of the fits. A notable distinction is that in the field of NS, an approach based on the theory of slow rotations and the Hartle-Thorne formalism can be used \cite{Hartle:1967he}, together with studies on total and rapid rotations. Generally, studies of slow rotations yield universal relations for NS and Quark stars with errors below $1\%$ \cite{Yagi:2013awa,Yagi:2014bxa,Yagi:2014qua,Adam:2020aza}, whereas studies involving total rotation typically show associated errors ranging from $1\%$ to a maximum of $20\%$ for some cases \cite{Yagi:2014bxa,Yagi:2016bkt,Doneva:2016xmf,Doneva:2017jop,Guedes:2024gxo}.

Taking the PHS together with the PS models, we have found that similar universal behavior exists between different vectorial star models. There is also an important point: the coincidence between the PS and the limiting case of PHS when $\alpha=10$ allows us to treat the set of PHS stars as a complete system without using the PS model separately. 

In the subsequent subsections, we demonstrate, on the one hand, the discrepancy between our vector and hybrid models with the boson star fittings, and additionally, we elucidate how these models occupy distinct yet close regions in the parameter space. On the other hand, we show the PS and PHS universal relations. In the conclusions section, we will explain how these observations contribute to a significant disruption in the potential for astrophysical observation to distinguish between two possible types of stars in an event, the similarities between the models and Kerr-BH, and also the importance of this kind of relations from a theoretical perspective.

 \subsection{I-$\chi$-Q multipoles space comparison}
 \label{ichiqsect}
With the improvements in the coefficient fittings, and also the addition of some models, reaching secondary branches in the mass-frequency curves, and using higher self-interaction constants for the quartic potential (all the used models are shown in  \cref{appendix}), it was shown in \cite{Adam:2023qxj} that the universal relations still hold with an error of less than $1\%$ for $n=1$ BS. We have obtained the moment of inertia as a function of the spin parameter and the quadrupole moment. The surface fitting used the expression 
\begin{equation}
    \begin{split}
       &\beta=A_0+A_s^m\chi^m\left(\alpha-B\right)^s,
    \end{split}
    \label{fit-func}
\end{equation}
where
\begin{equation}
  \begin{split}  
    &\beta=\sqrt[3]{\log_{10}{\bar{I}}},\\&
    \alpha=\log_{10}\bar{Q}
    \end{split}
    \label{fit-func-par}
\end{equation}
and $s={1,2,3}$, $m={0,1,2}$. Fitting coefficients are given for all the above surfaces in \cite{Adam:2023qxj} . 
It is noticeable how the difference between the fitted surface and the real data is always less than $1\%$ for $n=1$ BS, whereas results for PS are in strong disagreement with the BS fittings, showing that, although both objects are versions of bosonic stars, the universality between them is broken.

\begin{figure*}[]
\centering
\subfloat[]{%
  \includegraphics[clip,width=0.94\columnwidth]{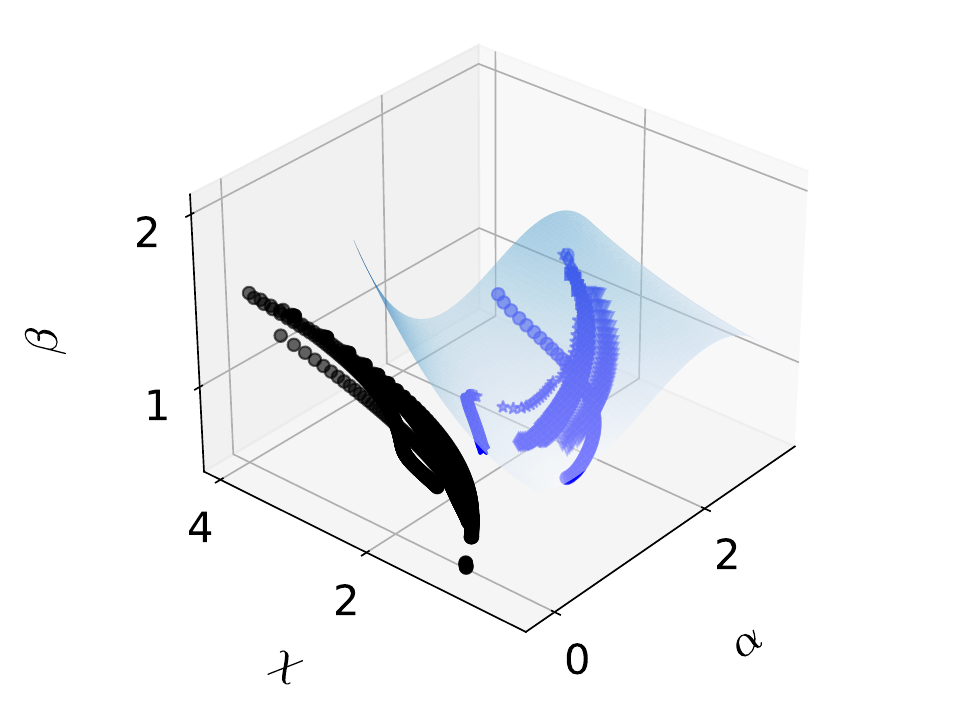}%
}

\subfloat[]{%
  \includegraphics[clip,width=0.94\columnwidth]{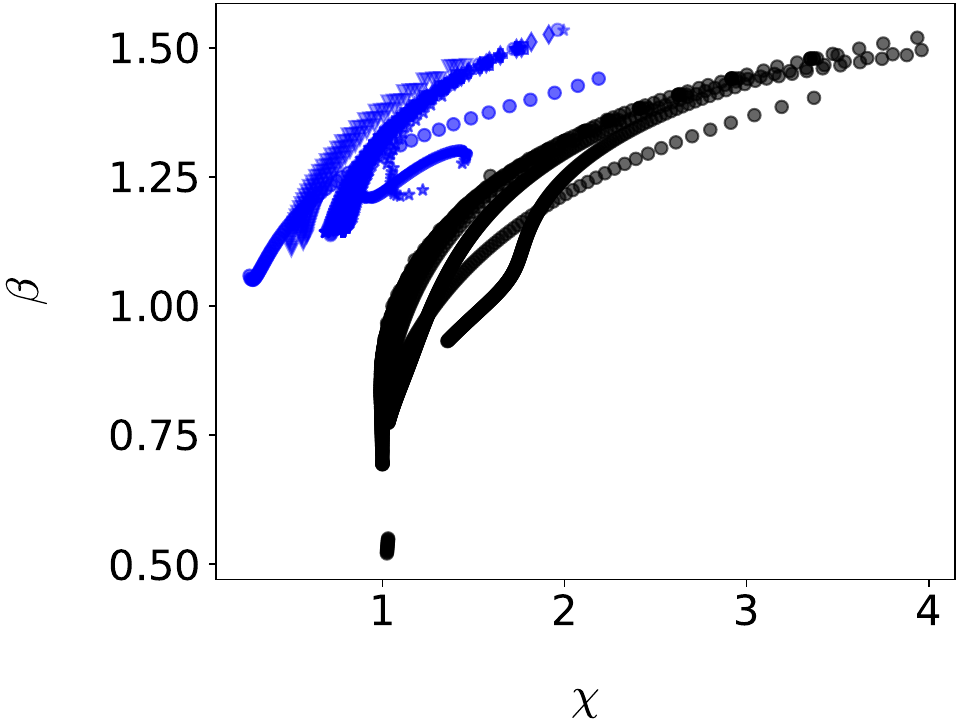}%
}

\subfloat[]{%
  \includegraphics[clip,width=0.94\columnwidth]{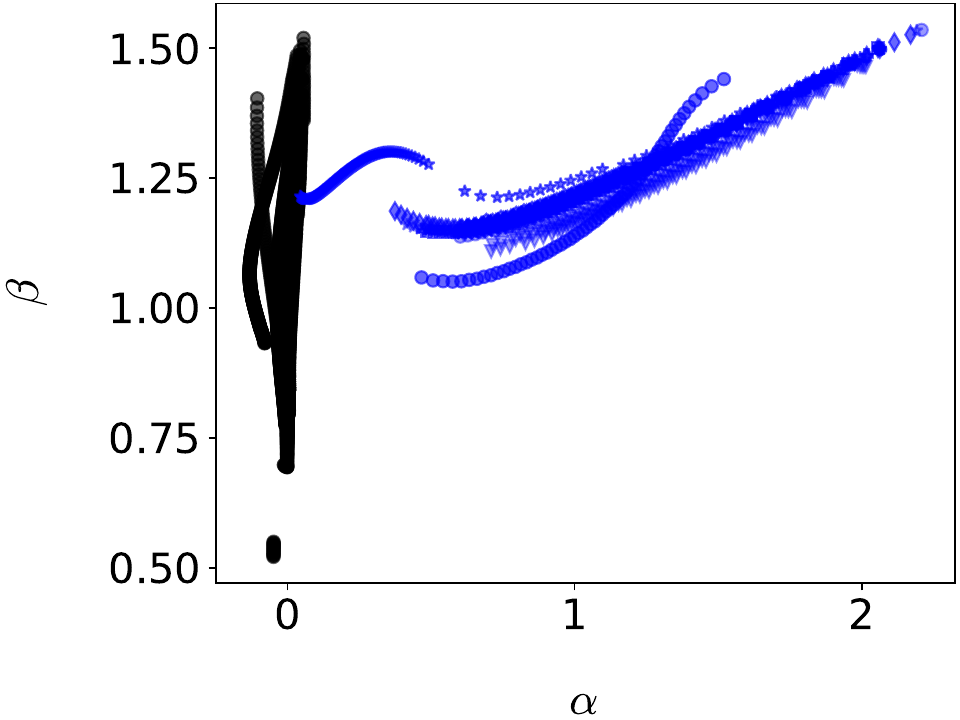}%
}

\caption{Universal $\beta-\chi-\alpha$ surface for $n=1$ spinning BS fitting the data points in blue, while $n=1$ vectorial boson stars are represented with black dots. It is clear that PS and PHS do not belong to the BS universal surface. The various scalar and vectorial models discussed here are referenced in \cref{appendix}.}
\label{figpsbs}
\end{figure*}

From \cref{figpsbs}  we can extract some important remarks.
As previously noted, the PHS models do not exhibit the same behavior across the 
$I-\chi-Q$ space than the scalar BS. It is impossible to say whether the vectorial star data depicts a fittable surface just by visual inspection.

Despite the differences, the complete dataset generally outlines a comparable region, illustrating that the multipoles for both scalar and vectorial stars are sufficiently similar to confirm shared astrophysical characteristics. Notably, quantities like the moment of inertia and the spin parameter align with those observed in BS, though PS models display a lower moment of inertia. The divergence in the quadrupolar moment is significantly less pronounced in vectorial models, suggesting that vector stars are generally less deformed and have lower inertial properties. Intriguingly, this could be linked to the atypical ground state of static PS models as suggested by \cite{Brito:2024biy}. Given the impossibility of continuous transitioning from static to spinning states in this bosonic framework, it is plausible to expect that, under certain conditions, a static PS might gain angular momentum. Then this initially non-spherical static state could, while rotated, become more spherical, resulting in a stronger object under deformation by rotation. Both star families occupy similar yet unique regions within the parameter space, highlighting their diverse structural and dynamic characteristics. However, while our BS data show universal behavior, vectorial stars deviate from this common fitting function.



\begin{figure*}[]
\subfloat{%
  \hspace*{0.9cm}\includegraphics[clip,width=1.0\columnwidth]{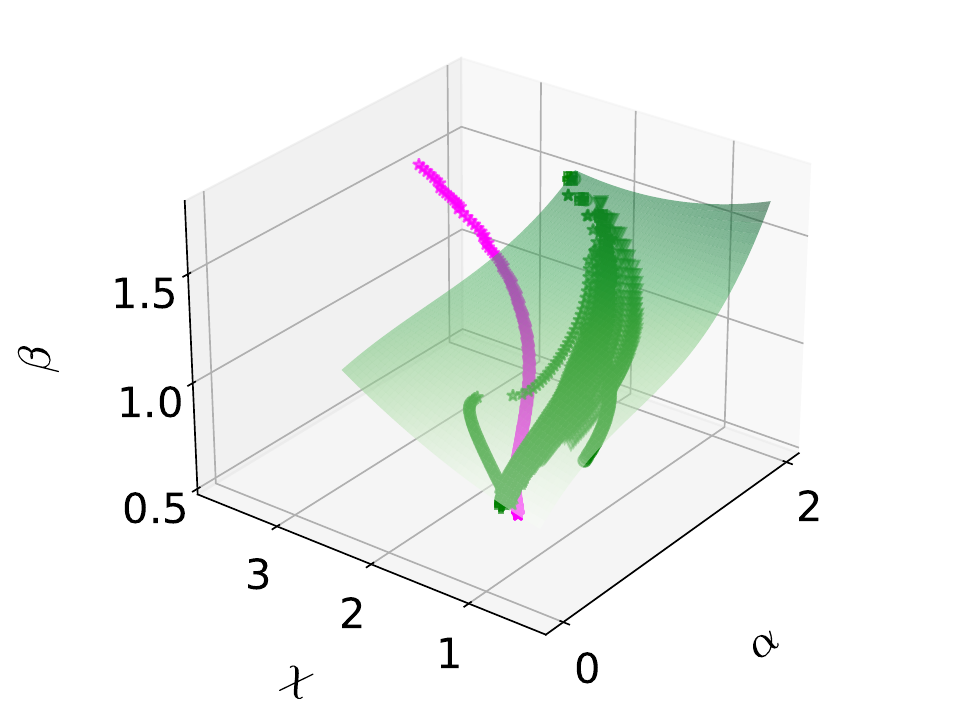}%
}

\subfloat{%
  \hspace*{0.9cm}\includegraphics[clip,width=0.94\columnwidth]{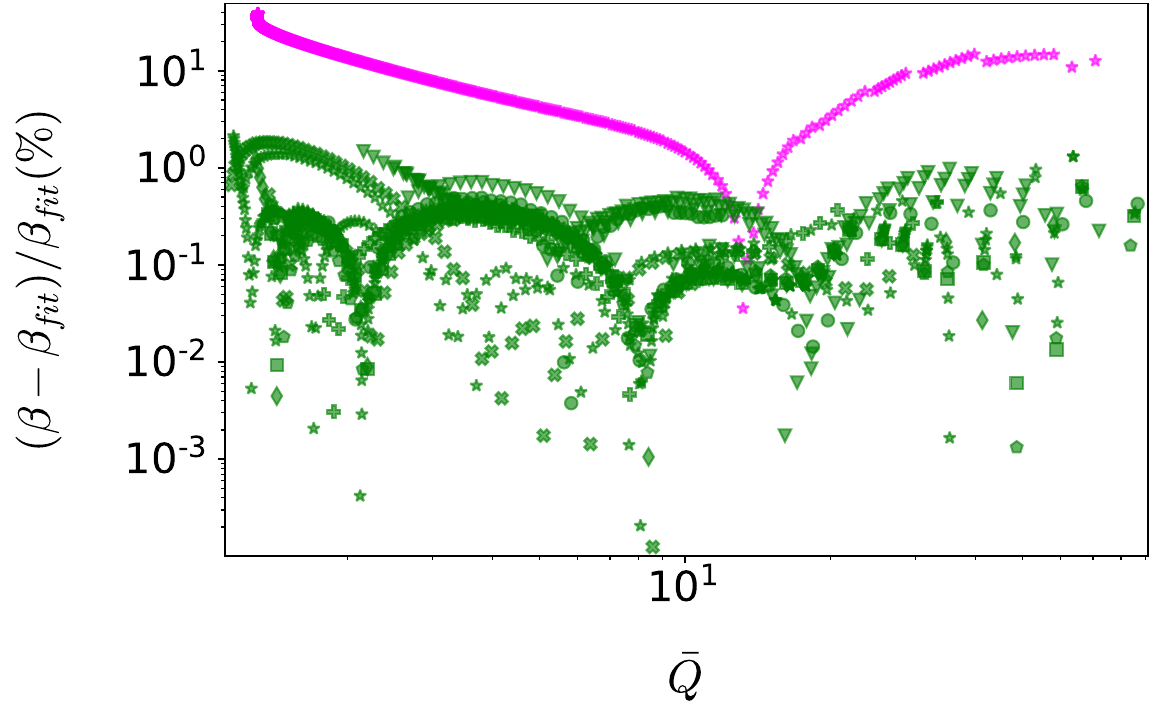}%
}
\caption{Universal $\beta-\chi-\alpha$ surface for $n=2$ spinning BS fitting the data points (all in green) and $n=2$  PS curve in magenta (upper panel). The relative difference between the data and fitted value for BS (in green) and the disruptive tendency for the PS, shown in magenta is observable from the lower panel. The deviation of PS data from the BS fit is evident overall, except for a small region where the multipole moments of vectorial stars exhibit proximity to the scalar counterparts. This observation implies the existence of a spectrum of solutions wherein both families display similarities within the $I-\chi-Q$ space. By comparing these data with other families of relations referenced in \cref{s3m2}, we note that the scalar adjustment aligns with the PS point across different $Q$ ranges, thereby allowing for the differentiation between scalar and vectorial $n=2$ stars. However, the underlying reason for the close proximity of these objects for certain multipole moments remains unknown.  }
\label{supern2}
\end{figure*}

For completeness, we have obtained the $n=2$ curve for the common PS. If we plot the second harmonic index stars in our $3D$ space, together with their BS counterparts \cref{supern2},  we can observe that they behave very similarly, making it much harder to distinguish in this scenario than for $n=1$ stars. This is particularly clear in the error plot (\cref{supern2} lower panel). We can observe both the relatively similar behavior and the remaining difference between the BS fitting and the PS data for $n=2$. The absence of $n=2$ PHS is primarily attributable to the numerical complexities involved in modeling these objects.

Looking at the lower panel of \cref{supern2}, we observe that although the data seem to be sufficiently close for a unique fitting, the errors range from $0.1\%$ to $300\%$  for the PS data, which means that vectorial boson stars are not in the same parameter space region.

Notably, the analysis of Proca stars for $n=3$ was not conducted owing to computational limitations. The higher excited state solutions for this harmonic index were not attainable with the same level of precision as achieved for the lower harmonic indices, which explains the exclusion of this case from our current study.

 \subsubsection{I-$\chi$-Q relations for vectorial stars}

\begin{figure*}[]
\subfloat{%
  \hspace*{0.3cm}\includegraphics[clip,width=1.0\columnwidth]{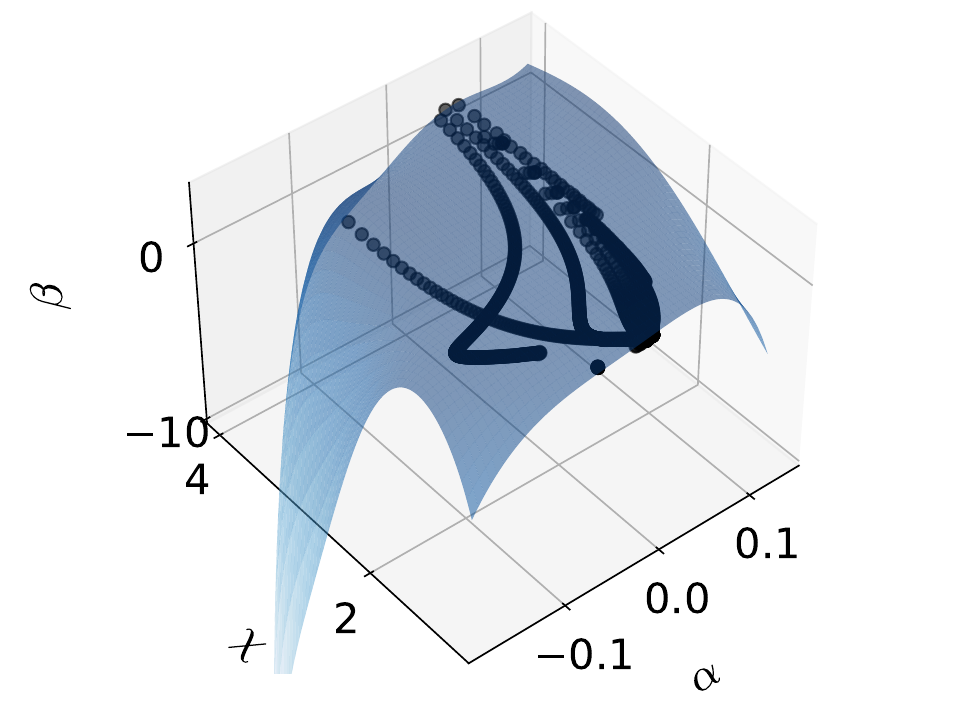}%
}

\subfloat{%
  \hspace*{0.3cm}\includegraphics[clip,width=0.94\columnwidth]{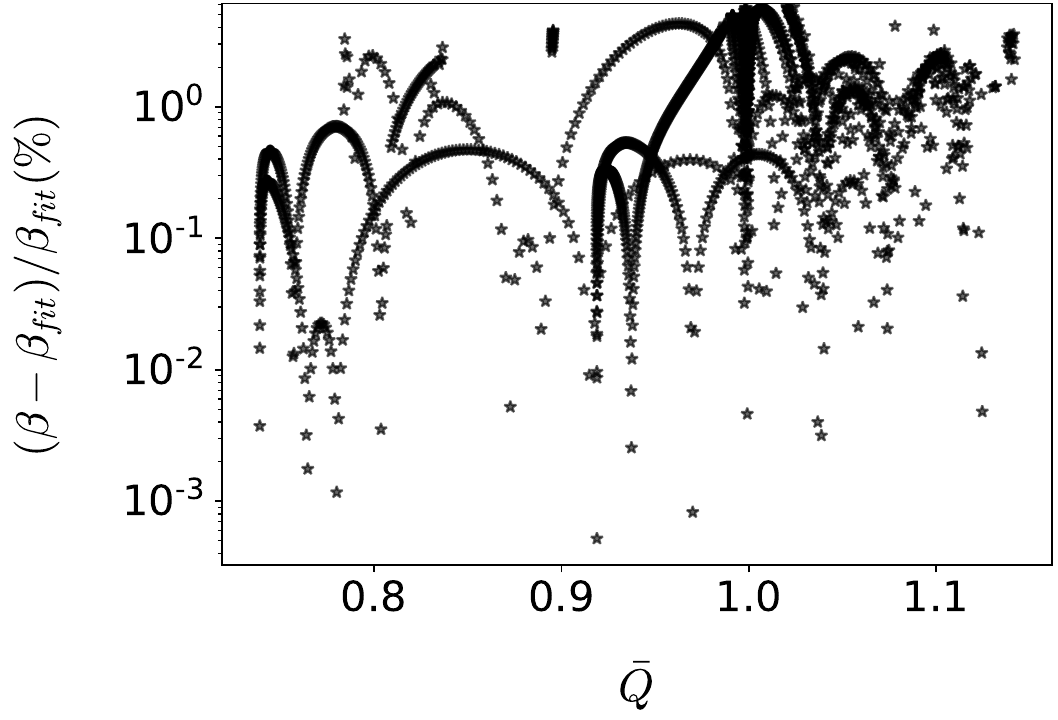}%
}
\caption{Universal $\beta-\chi-\alpha$ surface for $n=1$ spinning PH and PHS fitting the data points (upper panel). The relative difference between the data and the fitting surface is shown in the lower panel.   }
\label{superIXQP}
\end{figure*}

The treatment for the PS and PHS alone allows us to extract a universal fitting for these kinds of stars. Having the vectorial data in the $3D$ space \cref{superIXQP}, it is clear that all the black dots form a smooth surface that can be fitted through a procedure similar to the BS case. The kind of function used for this purpose is,
\begin{equation}
    \begin{split}
       &\beta=A_0+A_s^m\chi^m\left(\alpha-B\right)^s,
    \end{split} \label{fit-func-p} 
\end{equation} 
which is the same as \cref{fit-func} and $\alpha,\beta$ are \cref{fit-func-par}, but now with $s={1,2,3,4,5,6}$;  $m={0,1,2,3,4,5,6}$ and the condition $A_s^m=0$ if $m>s$ (the full table of coefficients is shown in \cref{appendixB}).

What we have found with respect to the universality mentioned before is that for both PS and PHS the moment of inertia, the quadrupolar moment, and the spin parameter $(\beta,\alpha,\chi)$ and, therefore, $(\bar{I},\bar{Q},\chi)$ are related in such a way that knowing two of them, the third is determined. The most usual way of showing this is just via the expression \cref{fit-func-p}, which equates $\beta$ with the logarithm of the reduced quadrupole moment and the spin parameter through the matrix of coefficients. 

Let us clarify our criteria for the choice of the fitting function and the coefficients matrix. Our criteria for the best fitting equation is the relative error between the proper data, i.e., each of the black dots populating \cref{superIXQP} which encode the multipolar information for a specific star, and the fitting value. In our case, as we are obtaining the function for $\beta$, the fitting value corresponding to a given star is what we call $\beta_{fit}$ which is the quantity obtained through a given fitting, with the associated numerical coefficients, and with the input of $\alpha$ and $\chi$ for each case. As shown in lower \cref{superIXQP}, the error between data and fitting for each case is obtained by
\begin{equation}
    Error=100\cdot\frac{|quantity-fitted\ quantity|}{fitted\ quantity}
    \label{error}
\end{equation}
The same procedure was used in \cite{Adam:2023qxj,Adam:2023qxj}, but we have a new issue regarding the degeneracy of surfaces. As we have less fitting parameters in usual BS universal surfaces, it is quite clear which surfaces minimize the errors. Now we have fitting surfaces that are more complicated, and the decision criterion for choosing a set of values for $s,m$ over other ones is harder. This means that while a minimal error argument is used for our selection for the fitting function and values, other choices would give similar errors. 

After seeking the most accurate choice of fitting function, we get a maximal error with a $~ 6\%$ deviation between the stars for which the fitting matches worst, corresponding to the region of $\Bar{Q}\sim 1$ and for the PHS models that are further from the PS limit. With the above error in the worst case, and having less than a $3\%$ in most cases, as shown in the lower plot of \cref{superIXQP}, we can ensure the existence of quasi-universal $I-\chi-Q$ relations for vectorial stars.

\subsection{Spin octupole}

In this work, we also consider higher-order multipoles. As \cite{Adam:2023qxj} did for spinning scalar BS, the spin-octupole moment is studied now for vector boson stars and compared with their scalar cousins. For the scalar case, some seminal work has been done by \cite{Ryan:1996nk,Vaglio:2022flq} in the strong coupling constant approximation.

Building upon our preceding investigations into universal behaviors, this study explores the potential for three-dimensional relationships among higher-order multipoles within the vectorial boson star paradigm. Furthermore, consistent with the approach in prior sections, we juxtapose these novel models against their scalar analogs. Our presentation will focus exclusively on the fitting for the spin octupole moment, understanding that the analysis of the mass hexadecapole scenario is comparable.

Firstly, we emphasize that we are not using the same definitions for the fitted spin-octupole as in \cite{Adam:2023qxj}, making the scalar BS error plot look worse in the present manuscript (the maximal error found in the aforementioned manuscript was smaller than the $3.5\%$ ).

\begin{figure*}[]
\subfloat{%
  \hspace*{1.0cm}\includegraphics[clip,width=1.0\columnwidth]{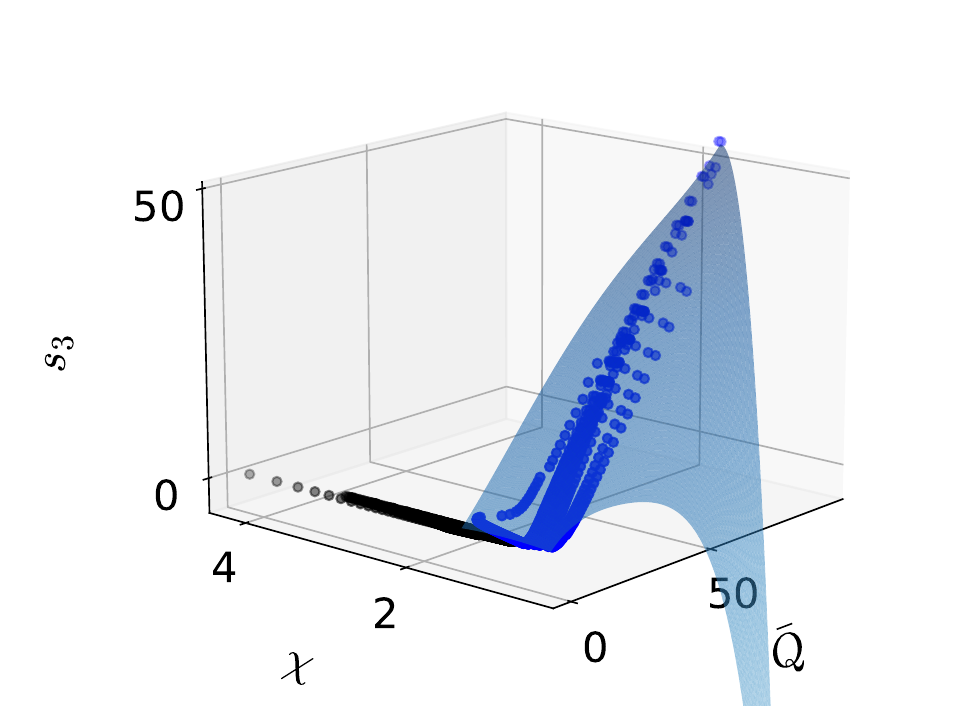}%
}

\subfloat{%
  \hspace*{1.0cm}\includegraphics[clip,width=0.94\columnwidth]{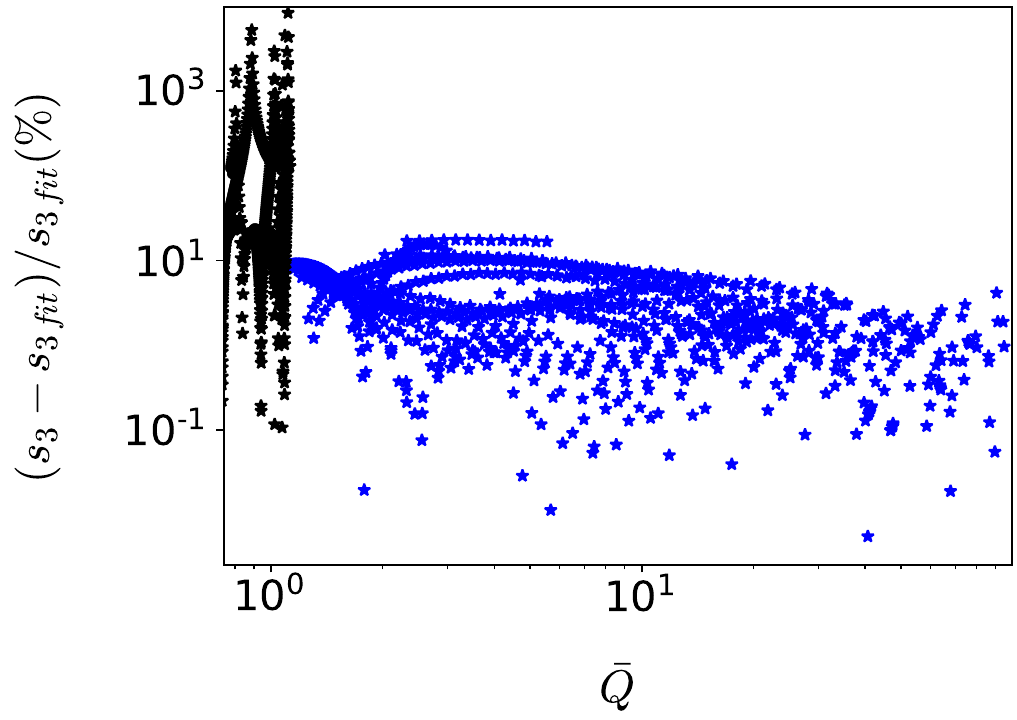}%
}
\caption{In the upper panel we have our data set in the $\chi-\Bar{Q}-s_3$ space, where blue dots and the blue surface describe the BS data and fitting, while black dots represent the vector stars. In the lower panel, we present the error plot, doing the same surface adjustment over the BS data, and showing not only the deviation from the surface for the BS (in blue) but also for the vectorial cases (in black).   }
\label{s3}
\end{figure*}

For stars with harmonic index $n=1$,  the upper panel in
\cref{s3} reveals the absence of absolute universal behavior for the spin octupole multipole between our vector star configurations and their scalar counterparts. In the three-dimensional diagram, the black dots congregate within a confined region, exhibiting similar behaviors to one another. Yet, they do not form a cohesive surface with the BS. Intriguingly, certain vectorial models align with the BS surface, rendering the differentiation between vectorial and scalar stars based on the $s_3$ multipole unfeasible in the highly relativistic domain. This observation is further elucidated by the lower panel in 
\cref{s3}, where a significant number of black dots demonstrate an error margin below $1\%$ when compared to the BS universal surface.

It is crucial not to lose sight of the overarching perspective of the analysis when interpreting this result. In a hypothetical astrophysical scenario, the dominance of the moment of inertia and the first multipoles would serve as a useful tool, breaking the degeneracy problem through the 
$I-\chi-Q$  relations. The lack of a more pronounced separation in the $s_3$ multipole space is theoretically interesting, suggesting that these different families of objects are not as divergent in their spacetime structure as it could be assumed from the $I-\chi-Q$ analysis.

\begin{figure*}[]
\subfloat{%
  \hspace*{0.0cm}\includegraphics[clip,width=1.0\columnwidth]{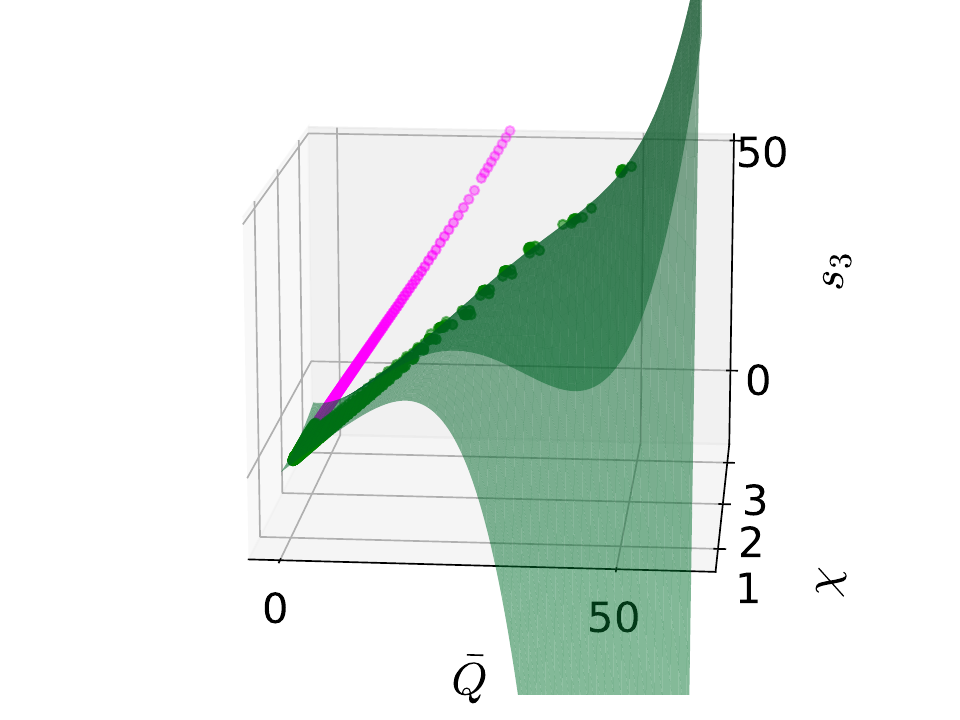}%
}

\subfloat{%
  \hspace*{0.0cm}\includegraphics[clip,width=0.94\columnwidth]{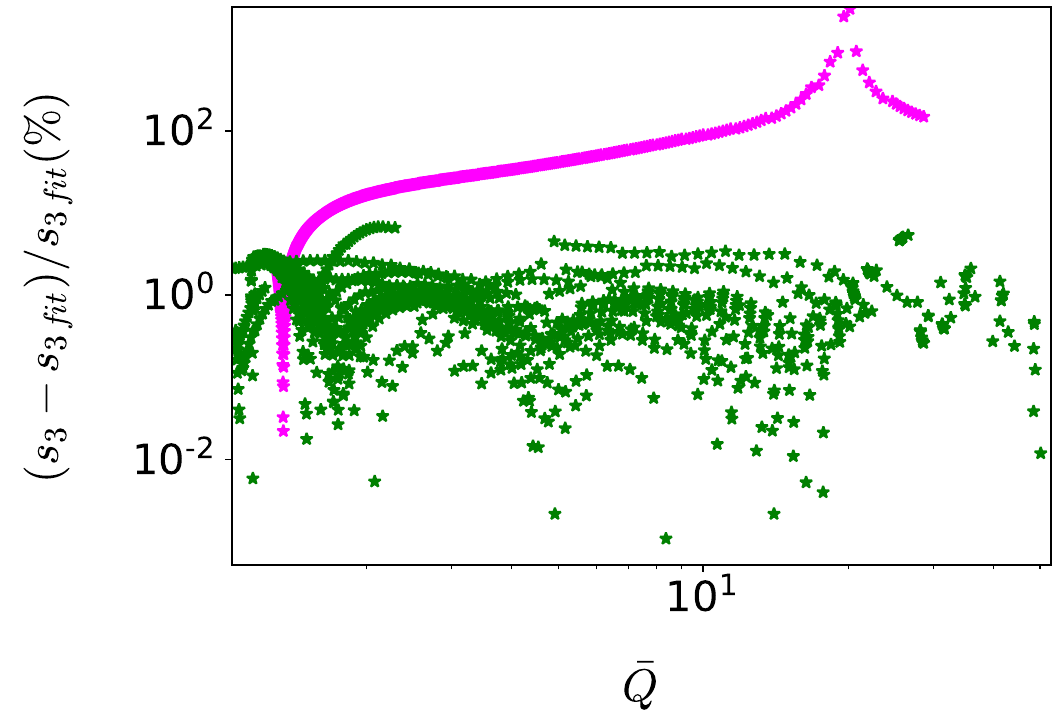}%
}
\caption{In the upper panel we show the $\chi-\Bar{Q}-s_3$ space for $ n=2$ BS (green) and $ n=2$ PS (magenta). The lower panel shows the error plot.}
\label{s3m2}
\end{figure*}

In extending our analysis to the 
$n=2$ dataset \cref{s3m2}, we reaffirmed the persistent divergence between PS and BS. However, akin to the observations made in the 
$I-\chi-Q$  analysis, our findings indicate a notably closer proximity between the datasets, with the PS models even intersecting the BS surface in the less relativistic regime. This is clear from the lower panel in \cref{s3m2} where we can see how many points have the same or less error concerning the BS fitting than other proper BS.

An exhaustive examination of the $m_4$ multipolar components is not presented because the outcomes and conclusions are very similar to those observed for the 
$s_3$ multipole. 

 \subsubsection{$s_3$-$\chi$-Q relations for vectorial stars}

\begin{figure*}[]
\subfloat{%
  \hspace*{0.3cm}\includegraphics[clip,width=1.0\columnwidth]{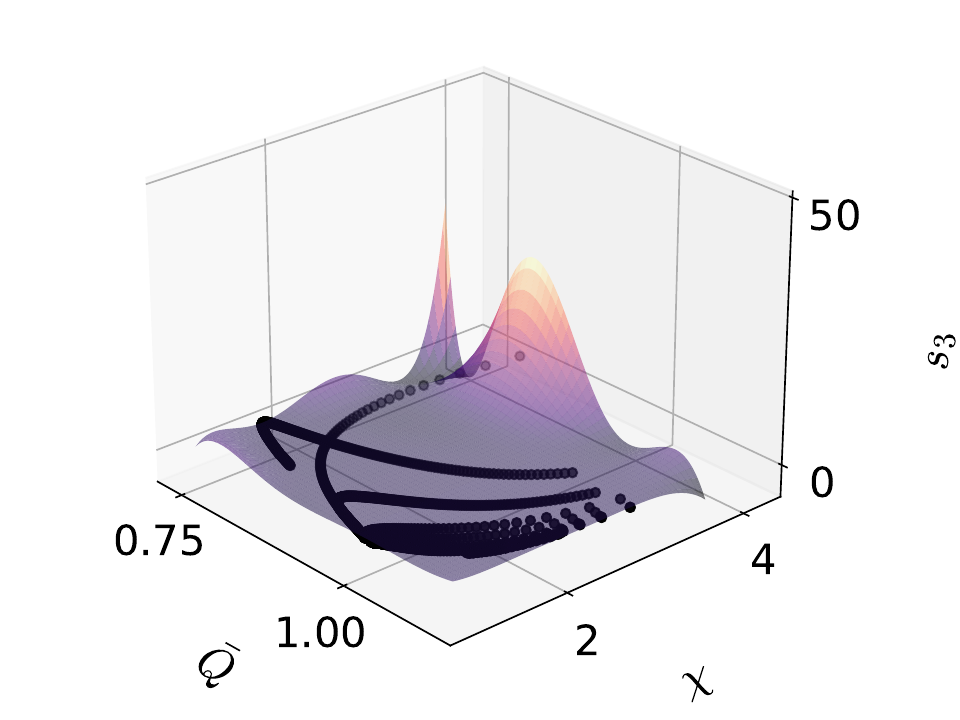}%
}

\subfloat{%
  \hspace*{0.3cm}\includegraphics[clip,width=0.94\columnwidth]{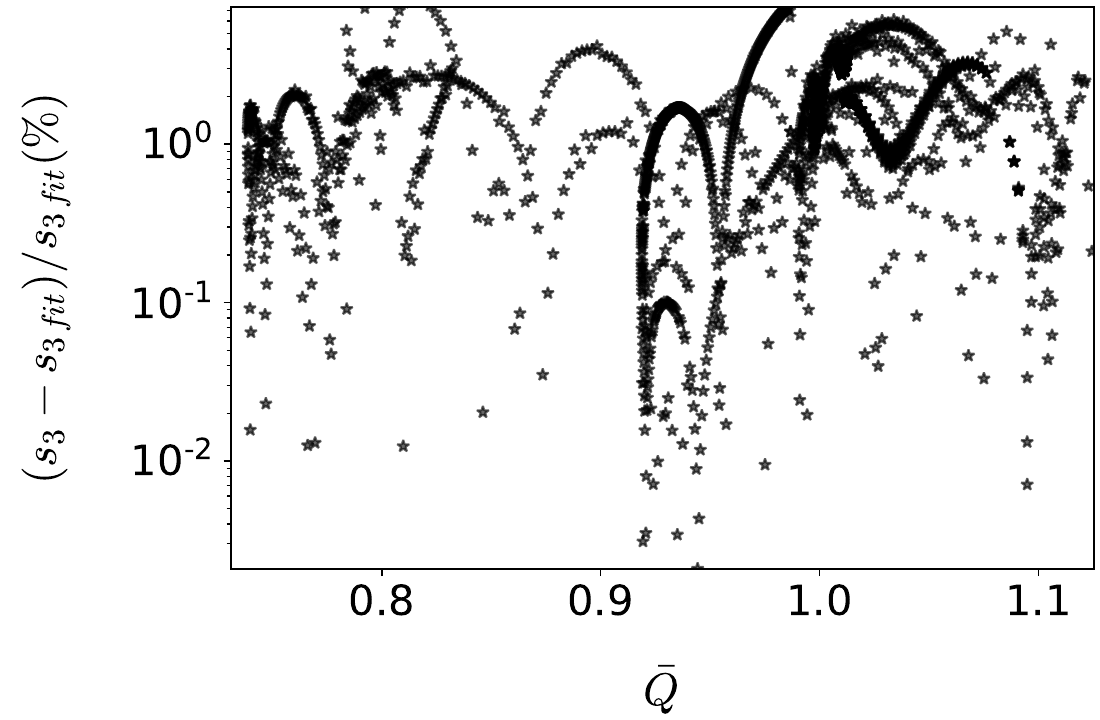}%
}
\caption{Universal $s_3-\chi-\alpha$ surface for $n=1$ spinning PS and PHS fitting the data points (upper panel). The relative difference between the data and the fitting surface is shown in the lower panel.   }
\label{supers3}
\end{figure*}

As we did in the previous subsection, the finding of a vectorial star universal or quasi-universal relation concerning the spin-octupolar moment can be done once data are treated separately. Taking all the PS and PHS as a set of vectorial data, we can easily see in \cref{supers3} that a surface is formed in the $s_3,\bar{Q},\chi$ space and the value for the spin-octupoles can be extracted through a fitting function and the pair $(\bar{Q},\chi)$.

The function that fits the data better, following the error criterion explained before, is the following,
\begin{equation}
   {s_3}=A_0+10^sA_s^m\chi^m\left(\Bar{Q}-B\right)^s, 
\end{equation}
where we now have $s=1,2,4,6$ and $m=0,1,2,3,4,5$. Look at the $s$ values, which are not consecutive and even except for $s=1$. 
A comment about the upper plot in \cref{supers3} and its peak shape in the small $\Bar{Q}$ high $\chi$ region must be made. As we do not have enough data in this limit, the fitting is less reliable, and the shape is not smooth. This must be taken as a numerical error that can be erased by having more models in this region. Some coefficients could change by expanding the data set in the mentioned parameter space. Still, the general result and the mathematical functionality of the fitting formula should not be affected, making our result general even with this issue.

Looking at the lower plot in \cref{supers3}, we can read that the maximal error obtained with the shown formulas is about $~7\%$, which allows us to ensure that there is a universal or quasi-universal relation between $s_3,\bar{Q}$ and $\chi$.


\subsection{Compactness}

\begin{figure*}[]
\centering
\subfloat[]{%
  \includegraphics[clip,width=0.94\columnwidth]{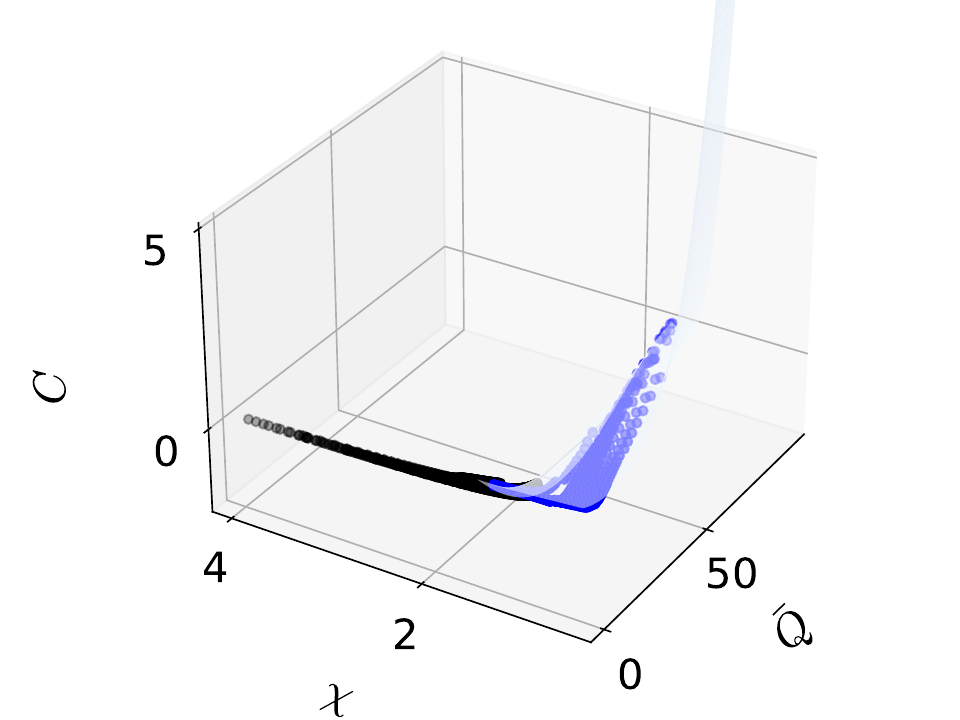}%
}

\subfloat[]{%
  \includegraphics[clip,width=0.94\columnwidth]{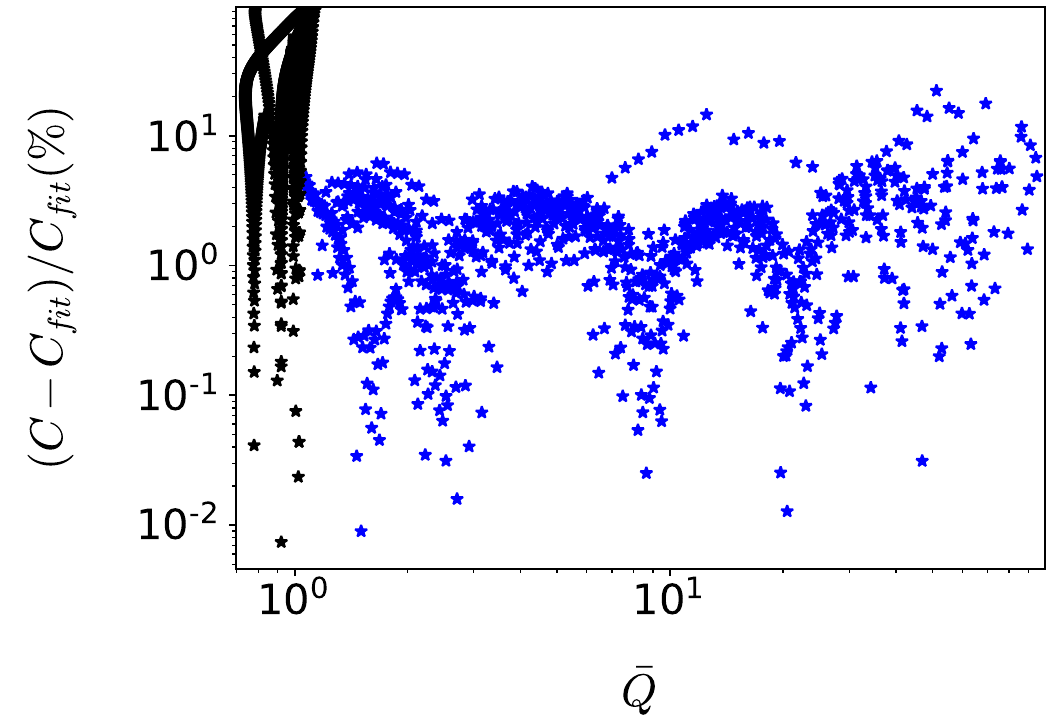}%
}

\subfloat[]{%
  \includegraphics[clip,width=0.94\columnwidth]{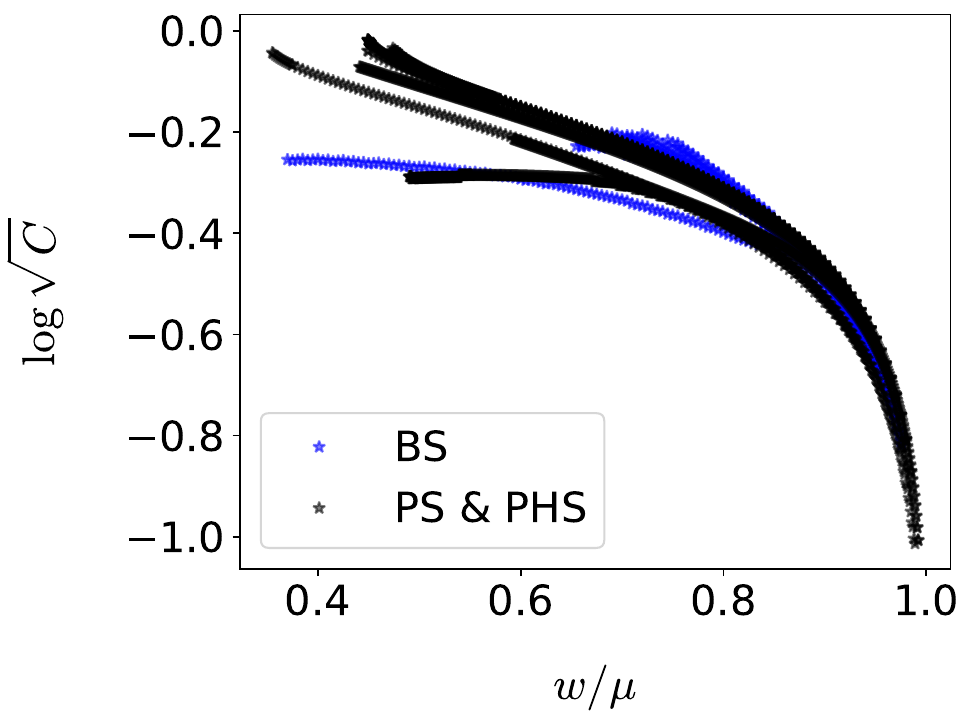}%
}

\caption{BS are represented in blue while PS and PHS are shown in black, all of them are $n=1$ states. The upper panel shows the full $C-{\chi}-{\Bar{Q}}$ space. The middle panel shows the error between the BS fitting and the full data set. In the lower panel, we show how the $\log{\sqrt{C}}$ behaves in front of the internal field frequency $w/\mu$, emphasizing the clear similarity between the shown compact stars.}
\label{comp}
\end{figure*}

 In \cref{comp} upper panel, we plot our data in the $\textit{C}-\log\Bar{\chi}-\log\bar{Q}$ parameter space. The blue dots and surface belong to the scalar BS; black points are our vector models. It is clear that the BS data shape a soft surface, which can be understood as a family of universal relations with a less than $2.5\%$ error for $n=1$ stars, as reported in \cite{Adam:2023qxj}. The space spanned by the vector stars occupies a small region because their quadrupolar momenta are always smaller than for BS.

The central panel analysis shows that vectorial and scalar stellar configurations exhibit distinct properties, in general, within the specified domain, rendering a unified model untenable. But as in the previous case for the $s_3$, there is a region where the vectorial and scalar data behave similarly. This overlap makes some scalar BS, PS, and PHS identical in compactness, spin parameter, and quadrupolar moment. Despite this, we could break the degeneracy just by comparing the errors between the scalar BS fitting and the vectorial universality, and we would check that the proper vectorial surface would be better. In addition, if a GW event is observed from an unknown source, the extrapolation of the angular momentum, mass, quadrupolar momentum and the size of the object would allow to break the degeneracy between scalar and vector boson stars just by identifying the position of the data in our $3D$ plot.

To further compare the compactness of these entities, a bi-dimensional plot was introduced to correlate another variable, the internal frequency 
$w/\mu$. The subsequent lower graph illustrates the relationship between the logarithm of the compactness' square root and 
$w/\mu$ by ensuring a consistent parameter range for both model types. This plot serves to clarify that, contrary to potential misinterpretations from the preliminary panel suggesting a disparity in this variable, the observed discrepancies are primarily attributable to the reduced quadrupolar moment in vectorial stars, which skews the perception of comparability amongst these compact objects.

The investigation was extended to 
$n=2$ stellar configurations focusing on compactness yielding the following summary. In the context of boson stars characterized by a higher harmonic index, encompassing both Proca and Scalar types, the proximity between the respective families increased. However, the vectorial stars remained outside the viable fitting region, with a majority of PS and PHS instances exhibiting an error margin of approximately around 
$60\%$. This indicates the impracticality of applying a singular-fitting approach to the entire dataset.

\subsubsection{Compactness relations}

In the compactness case, and doing the same procedure of analyzing the set of vectorial star data just by itself, without taking into account the scalar stars, we have found the lesser errors using the following surface,

\begin{equation}
   C=A_0+A_s^m\chi^m\left(\Bar{Q}-B\right)^s,
  \label{fitm4} 
\end{equation}
with $s=0,1,2,3,4$ and $m=0,1,2,4,6,8$.

\begin{figure*}[]
\subfloat{%
  \hspace*{0.0cm}\includegraphics[clip,width=1.0\columnwidth]{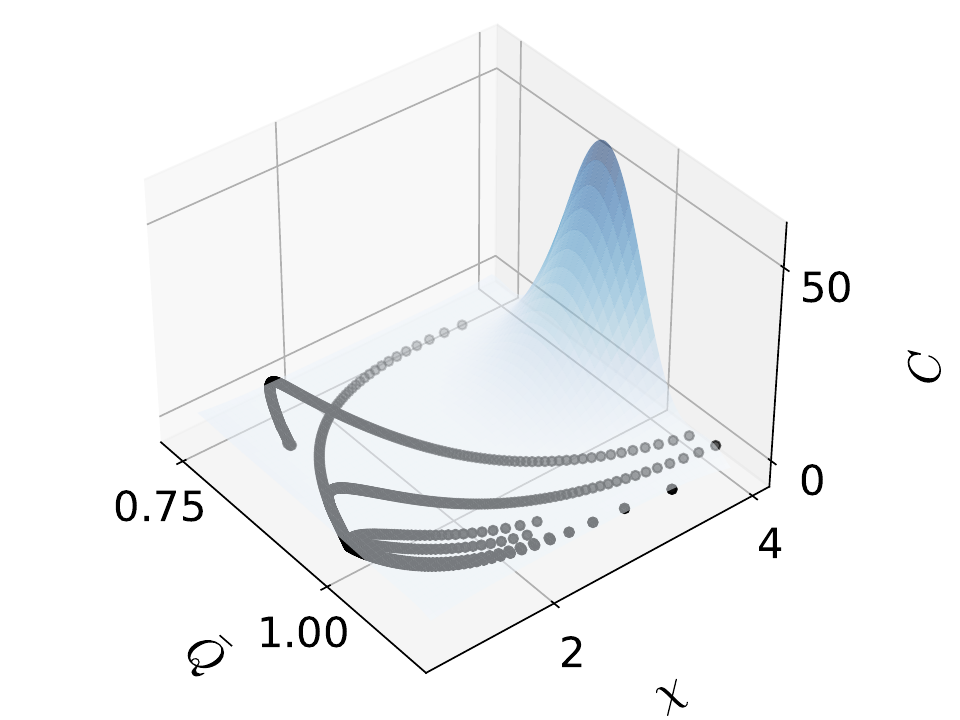}%
}

\subfloat{%
  \hspace*{0.0cm}\includegraphics[clip,width=0.94\columnwidth]{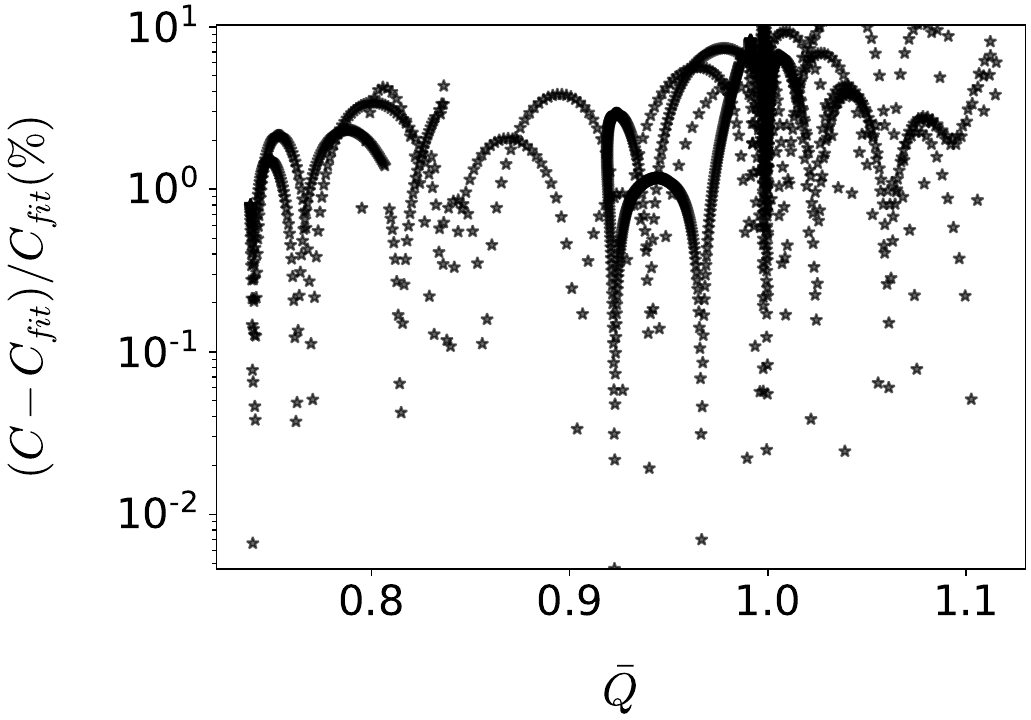}%
}
\caption{Universal $\mathcal{C}-\chi-\alpha$ surface for $n=1$ spinning PHS fitting the data points (upper panel). The relative difference between the data and the fitting surface is shown in the lower panel.   }
\label{superIXQ}
\end{figure*}

With the set of coefficients shown in \cref{appendixB}, the maximal error between data and fitting was $\sim 11\%$, which led us to ensure the existence of a universal or quasi-universal relation for the PS and PHS in the context of $\mathcal{C}$,$\bar{Q},\chi$ parameter space.

The role of compactness in the understanding of Universal relations is primordial; from the historical point of view, it was the first kind of relation found, and from modern perspectives, it is, for many authors, the better way of understanding this kind of universality. As highly compact objects tend to resemble black holes in their gravitational field, this could be an effective no-hair behavior that screens the importance of the matter properties in this very compact object, requiring fewer multipoles than expected to describe the space-time, as it happens for proper BH with the mass, charge and angular momentum.


\subsection{Comparison with Kerr multipoles and light rings}
\label{compkerr}

Bosonic stars, as mentioned, are often described as BH mimickers. This is best understood by considering the diverse multipolar behaviors of these exotic compact objects (ECOS), which approach the Kerr limit in specific models and under certain conditions. In the study of highly compact bosonic stars, it was observed that the quadrupole moment, the spin-octupolar moment, and higher-order moments converge toward unity—matching the Kerr values for parameters including mass and angular momentum. The above has a deep impact in GW science, as it would be a useful tool in the template matching and would also be a tool in constraining the ECOS \cite{Saini:2023gaw}.  This similarity means that a distant observer would find a black hole and a bosonic star appear remarkably alike. The Hairy Kerr Black Holes framework provides a natural method for validating these observations \cite{Herdeiro:2015gia,Herdeiro:2016tmi}.

\begin{figure*}[]
\subfloat{%
  \hspace*{0.25cm}\includegraphics[clip,width=1\columnwidth]{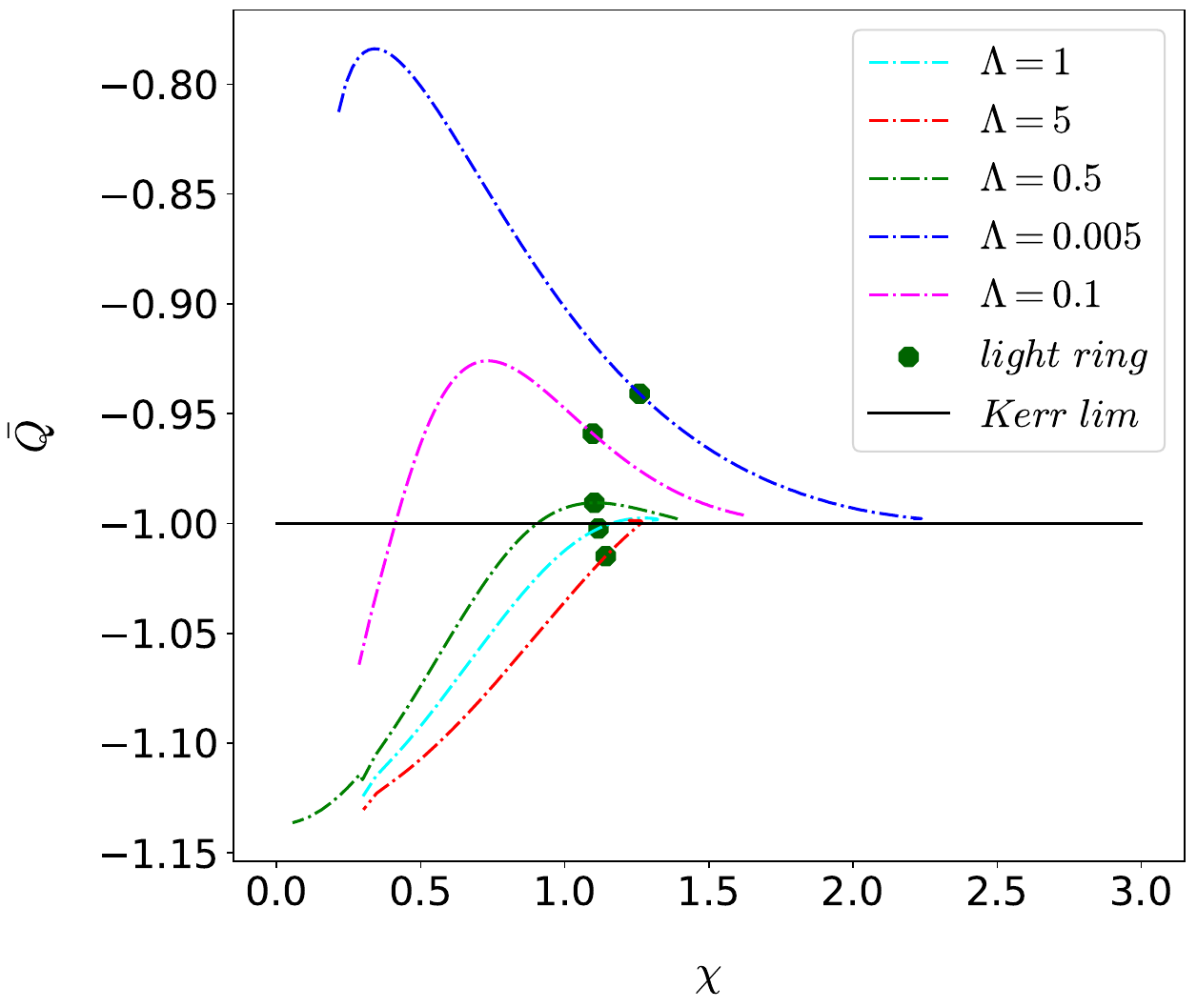}%
}

\subfloat{%
  \hspace*{0.25cm}\includegraphics[clip,width=1\columnwidth]{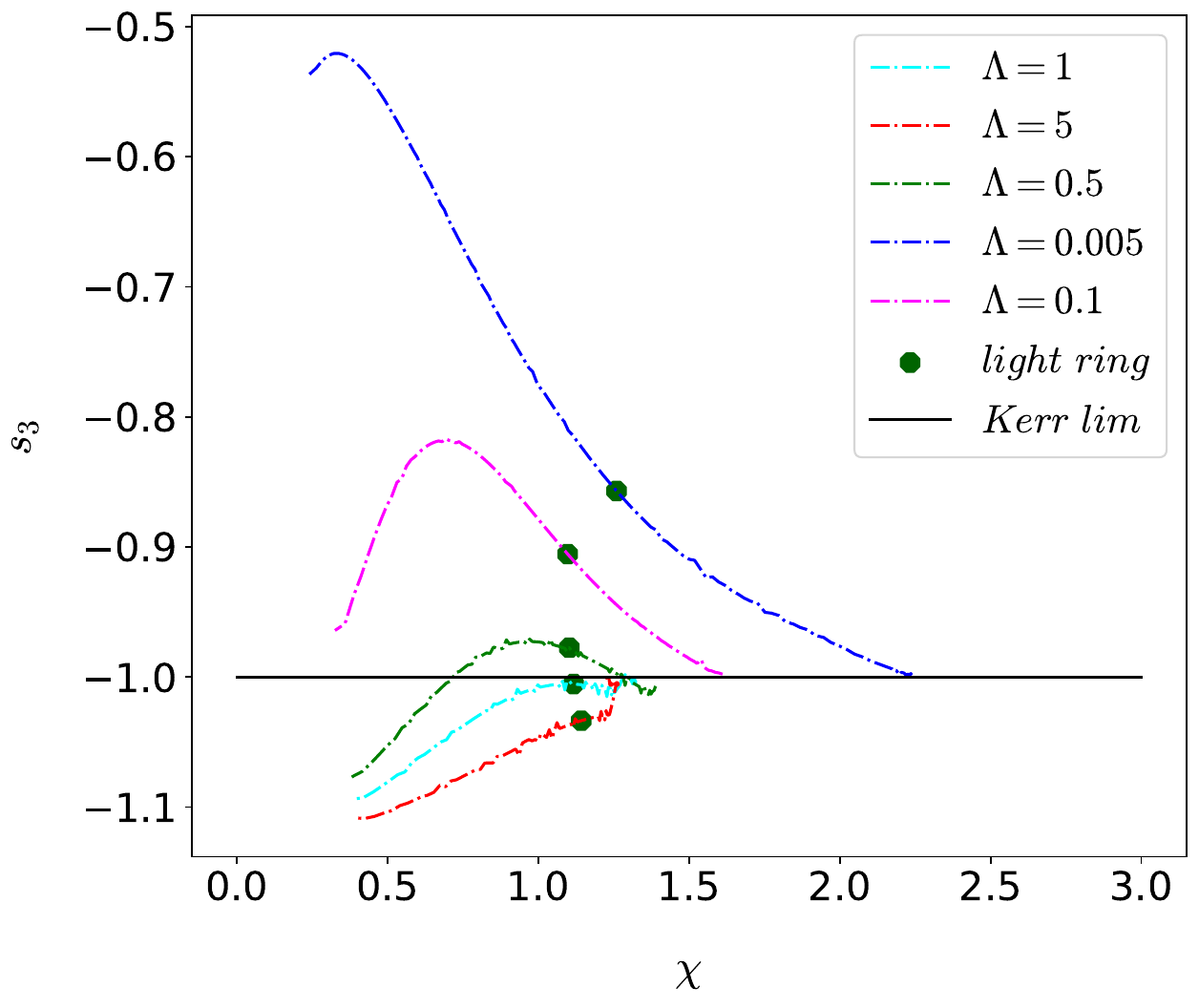}%
}
\caption{
We display five PHS models, each identified by its coupling constant $\Lambda$. We present numerous solutions corresponding to the stars associated with the frequencies $w/\mu$, so the different curves are formed by the sets of distinct solutions associated with the different frequencies for each coupling constant. These models demonstrate a clear convergence to the Kerr limit case $\bar{Q}=-1$, particularly for solutions with greater compactness. 
Additionally, the plot indicates the Light Ring (LR) threshold for each model, with stars displaying LR from that threshold to the highest value in the direction of increasing $\chi$. The lower plot presents a similar analysis for the spin-octupolar moment, demonstrating the alignment between the two multipoles
}
\label{lightrings}
\end{figure*}

This limit becomes particularly relevant when examining universal relations between multipoles. In certain areas of our $3D$ multipolar spaces for spin-octupole moments, the surfaces contract to a line, indicating that both the quadrupolar and spin-octupolar moments are approaching constant values, converging towards the Kerr limit. In the upper plot of \cref{lightrings}, we observe for some PHS models that the values of the quadrupolar moment approach $-1$, aligning with the Kerr limit. The lower plot shows a similar trend for the spin-octupolar moment, reinforcing the idea that, in the realm of very compact solutions when the values of $\omega/\mu$ are far from unity \cref{mbS}, horizonless compact objects like PHS can closely mimic the behavior of a black hole.


In the same line and due to the closeness in terms of compactness, some other spacetime features usually attributed to BH can occur for ECOS, namely light rings (LR) \cite{Cunha:2020azh}. A LR is an extreme case of light deflection in which the path followed or drawn by the light is closed over itself, a special kind of null orbit that is bound.
LR play a pivotal role in the observed appearance of black holes. They are intimately connected with the first-ever image of a black hole by the Event Horizon Telescope \cite{EventHorizonTelescope:2019ths}. This achievement significantly contributed to verifying some of the most remarkable predictions of Einstein's General Relativity in recent decades.
It was also shown that the ringdown involving any BH mimicker is directly affected by the  LR structures, which play a central role in the GW formation \cite{Cardoso:2016rao,Cardoso:2016oxy,Cardoso:2017cqb}. They found that the GW emitted shortly after an extreme-mass-ratio merger of a binary BH mimicker has consistent characteristics akin to those of a black hole ringdown, regardless of its internal nature. However, deviations from this pattern occur after a light-crossing time within the mimicker's interior, manifesting burst-like echoes in the signal.

In the case of topologically trivial, asymptotically flat ECOS, such as bosonic stars, LR come in pairs, under generic conditions, with one of the bound photon orbits being stable~\cite{Cunha:2017qtt,DiFilippo:2024mnc}. This implies a possible source of spacetime instability~\cite{Keir:2014oka,Cardoso:2014sna,Cunha:2022gde}. As such the understanding of LR in ECOS is an important diagnosis in the context of their physical viability. Note that this is not true for BH spacetimes, where a single LR may exist~\cite{Cunha:2020azh,Cunha:2024ajc} or for topologically non-trivial ECOS, such as wormholes~\cite{Xavier:2024iwr}.

A nonlinear treatment of the merger and ringdown for spinning scalar BS  was done very recently \cite{Siemonsen:2024snb}. The BH mimicker with stable LR obtained as a remnant aligns with our results in the sense of having objects with very similar multipolar structures but with internal matter degrees of freedom.


\subsection{Comparison and comments concerning other compact objects}

As discussed, e.g., in \cite{Adam:2023qxj}, neutron stars allow to independently adjust mass and spin parameter $\chi$, whereas, for both scalar and vector bosonic models, the variation of one parameter inherently dictates the other. Consequently, different BS and PHS models form universal surfaces in spaces of three observables for arbitrary frequencies, whereas this is true for NS only for a fixed frequency.
In other words, NS form three-dimensional hypersurfaces in four-dimensional parameter spaces where the frequency is one of these parameters.

\begin{figure*}[]
\centering
\hspace*{-0.90cm}\includegraphics[width=0.60\textwidth]{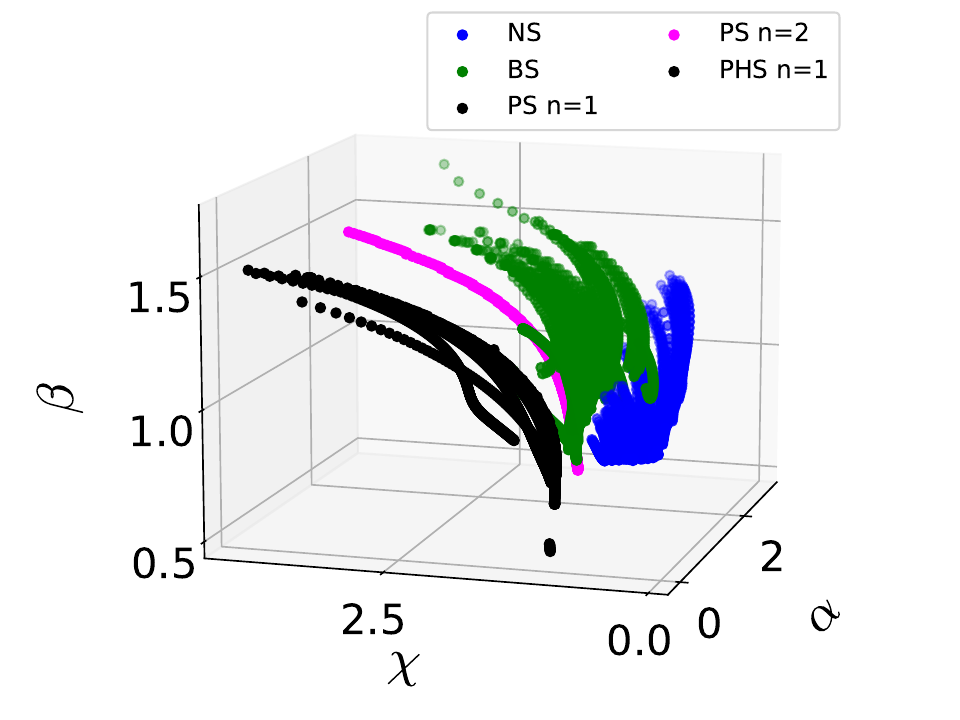}
\caption{Green dots correspond to $n=1,2,3$ BS data. Black points are PS $n=1$ data, and magenta dots are $n=2$ PS. Blue points are NS for different frequencies and several EOS, namely BCPM \cite{Sharma:2015bna}, AGHV \cite{Adam:2020yfv}, BPAL \cite{Zuo:1999vcl}, RNS-FPS \cite{Engvik:1994tj}, RNS-A \cite{Arnett:1977czg} and SLy \cite{Douchin:2001sv}. }
\label{NS}
\end{figure*}

In \cref{NS}, blue dots are rotating NS for various EOS ranging from low to high, mass shedding velocities, obtained using the RNS package  \cite{stergioulas1992rotating}. Green dots are $n=1,2,3$ spinning BS. Black dots are $n=1$ spinning PS and PHS, and magenta dots are $n=2$ PS.  Revisiting the previously discussed 
$I-\chi-Q$ features (where $\beta=\sqrt[3]{\log_{10}\Bar{I}}$ and $\alpha=\log_{10}\Bar{Q}$), it becomes apparent that a diverse array of compact objects exhibits remarkably similar behavior. While certain differences can be appreciated, it is evident that the analyzed stars possess shared characteristics. This implies that the space-times produced by these objects are fundamentally similar. 

As shown in \cite{Pappas:2013naa,Yagi:2014bxa}, the NS moments of inertia, spin parameter, and quadrupole moments fulfill some universal relations, and different BS share the same fate \cite{Adam:2023qxj}.  For our current set of vector stars, a universal fitting encloses all dots under the surface with an error of less than $6\%$. And many other features of interest can be extracted from \cref{NS}. First of all, it is clear from the above that $n=2$ Proca and Scalar Boson stars are closer between them than their fundamental states among themselves. The separation between the $n=1$ vector stars and the other compact objects is also noticeable, and the interesting result here is that PS and PHS have less quadrupolar momentum than NS and BS.
This observation is surprising, yet the comparison between BS and PHS is straightforward when considering their distinct field distributions in space-time. In contrast, with nearly spherical symmetry, NS and PHS diverge significantly in their quadrupolar moments. As discussed earlier in \cref{ichiqsect}, this disparity may stem from the considerable differences in the stationary states of each family. Alternatively, it could be that the internal forces within PHS are stronger than those in NS, leading to matter distributions in PHS that are less prone to deformation by rotation.

This could enclose fundamental information both in the GR context and particle physics. The quadrupolar momentum for a spinning object imprints the information about the deformation due to rotation, implying that higher momenta are present for more deformed objects.  Although we have comparable $\chi$ and $I$ for all the data set, the strong disagreement in the $Q$ is telling us directly that the compact objects resulting from vector bosons coupled to gravity are always much more resistant against rotational deformations than stars made of scalar bosons or fermions.

\subsection{Universality sources}

Ever since the discovery of universal relations like the $I$-Love-$Q$ relations \cite{Yagi:2013awa}, attempts have been made both for their mathematical derivation and for a better understanding of their physical origins \cite{Yagi:2014qua,Yagi:2014bxa,Yagi:2016bkt,Yagi:2016ejg,Guedes:2024gxo}. Here we advocate for two possible sources of the appearance of universal relations. Each of them has a rather different physical as well as mathematical origin.

First, compact stars are rather dense objects with a very strong gravitational field, which makes them similar to black holes. Therefore, {\it universal relations may originate in the no-hair property of black holes}. In other words, the significant compactness of the stars being analyzed here causes the surrounding space-time to be highly curved, and some universal behaviors might represent how the curvature masks details of the internal structure and the dynamics of the matter.

In the analysis of section three, referred to as $s_3$ in \cref{s3}, and the compactness discussed in \cref{comp}, it is clear that the most compact models of BS tend to align with PHS, consistent with the Kerr limit. This convergence trend across the two types of stars becomes more apparent in models with increased compactness. Notably, when evaluated with PHS data, the BS error function shows a strong fit only in the limit cases and only concerning the highest order multipole we consider and the compactness, as mentioned. This convergence is primarily attributed to the high curvature present in some models, which effaces the matter structure. This makes a distant observer to \textit{see} extremely compact BS and PHS (even NS) look like a BH in terms of the multipoles \cref{compkerr}. It is important to remember that those are defined at the limit $r\rightarrow\infty$ . Conversely, a universal behavior is still observed for stars that deviate significantly from the Kerr limit. However, the underlying reasons for this universality remain ambiguous from the current perspective. Although the universal surfaces persist in less compact models, they do not conform to the Kerr limit for the multipoles. This observation suggests that combining different factors might better explain the universality.

In scenarios where curvature is not particularly high, the distribution of matter or energy density becomes a critical factor. Different types of stars, NS, PS, or BS, exhibit distinct distributions, resulting in non-identical spacetime environments around them. This variation might explain why the universality surfaces generally differ despite having objects with similar or identical masses, angular momenta, and sizes. However, as observed earlier, these surfaces can converge in the case of limit models, especially for higher-order multipoles.

In the Newtonian case, a static perfect sphere only possesses the monopole moment. However, inducing rotation to such a perfect sphere does not lead to a change in its multipolar structure if this rotation does not induce deformations to the sphere. In the gravitation context, the perfect sphere is exemplified by the Schwarzschild solution. Again, the multipolar expansion possesses only the monopole term. On the other hand, rotation here will induce deformation in the source. The Kerr solution has an oblate shape due to the rotation. All multipoles of Kerr are present in the multipolar expansion and they vanish when setting the angular momentum to zero. They are spin-induced \cite{Poisson_Will_2014}.

Multipole moments quantify the decay of the gravitational field with distance, specifically capturing decay rates that scale as negative powers of $r$. Analyticity has been proven for  stationary, vacuum solutions of Einstein’s equations \cite{Hagen_1970,10.1063/1.525148}. We are considering bosonic stars in which the matter decays exponentially. Therefore, there exists an effective radius, such that, beyond such radius, the spacetime is in vacuum. The multipolar structure of those stars will, therefore, measure the deformation of this effective surface from a sphere.

Those stars admit a spherical symmetric solution. Once more, the multipolar structure is only given by the monopole. The rotating solutions can be conceived as the generalization of the spherical stars but with angular momentum. Therefore, their multipolar structure is, again, spin-induced. But why, then, when talking about universal relations, do we talk about a surface and not a simple line? Unlike black holes, where we do not have information about their interior and where the deformation on the horizon is only due to the spin, the rotating stars are not shells. Rather, the field will exert different pressures. The nature of the field will obviously have implications on the pressures it produces. Therefore, this deformation in the mass distribution must be taken into account, and therefore, we need an extra parameter, e.g., the momentum of inertia.

Secondly, there has recently been identified another possible source of universal relations. On the contrary to the previous one, it exploits a {\it non-gravitating limit of the matter field composing a compact object}. This is related to the recently found integral identities for non-gravitating nonlinear field theories supporting static or stationary finite energy solutions like solitons or $Q$-balls \cite{Adam:2024cem}. These identities give rise to some exact universal relations in the case without gravity. On the other hand, there are strong indications that under certain circumstances (asymptotic flatness of space-time and the existence of a global foliation with space-like hypersurfaces) the construction of these integral identities can be generalised to the case of self-gravitating field theories. This opens the intriguing possibility that at least some of the universal relations (or some combinations thereof) can be derived from exact identities which any static or stationary self-gravitating solution has to obey. This problem will be investigated in detail in the future.


\section{Conclusions}
\label{conclusions}

We investigated the existence of approximate, model-independent, universal, effective-no-hair relations between multipole moments and principal observables for BS, PS, and PHS of various winding numbers, finding the existence of the same kind of universal relations, presented for  BS \cite{Adam:2023qxj}. 
This kind of universality appears
for the vectorial cases we studied, including Proca Stars and Proca-Higgs Stars. Our results give us a possible method to break the degeneracy between vectorial and scalar boson stars in an astrophysical or GW signal scenario, because it follows from our multipole results that these compact objects can be easily distributed in distinct and clearly separated surfaces. Depending on the multipolar moments, we can distinguish between Proca, Boson, and Neutron Stars of similar mass and radius.

Our moment of inertia calculations for spinning vectorial stars also represent an interesting result by itself, making this paper the first to show such a result.

For horizonless entities, relations analogous to the exact no-hair theorem for black holes—termed universal or effective-no-hair relations—enable the determination of the external gravitational field with high precision using a finite set of multipole moments. This implies that, despite the presence of matter, it is not necessary to employ an infinite series of multipoles to describe the gravitating system within a suitable approximation. Such kinds of universalities extend to our vectorial dataset, making the vectorial stars enter the family of compact objects in which the universal behavior plays a principal role.

 Our PHS models are hybrid, coupling a scalar field to the vectorial star, making universal relations more remarkable. This universality underscores the importance of compactness effacing the impact (up to a certain point) of the matter composition in determining the physical characteristics of such entities. Additionally, our research indicates that certain models exhibit extremely compact configurations of PHS, which closely resemble or mimic the multipolar behavior of Kerr black holes. Moreover, these models also support the existence of closed light orbits, or light rings, up to a specific threshold. Although this supports the argument that these configurations effectively mimic Kerr black holes, it's important to remember that stability studies suggest that these highly compact models may have significant stability issues, casting doubt on their astrophysical reality. Consequently, there is a need to strike a balance: on the one hand, the high compactness achieved in a stable manner aligns with the non-hair theorems for Kerr black holes, suggesting their potential as BH mimickers. On the other hand, most ultra-compact models exhibit stability problems, necessitating caution when proposing these objects as BH mimickers.


The existence of universal relations has practical applications. Independent measurements for different linked quantities would allow inferring the related parameter. It is not only a powerful tool to deduce physical magnitudes but also serves to check some of them with the measurements given by GWs science. We have found universal or quasi-universal relations in our vector boson star scenario, so we can use them in the common manner.

We can also go further and use our results to guess in which region of the parameter space the third discordant quantity can appear, and use the GW results as checks for them.
Considering the significant recent advancements in GW observations, the implemented data analysis methodology could assume a pivotal role across various domains within the field. This is particularly true for predicting the most likely type of astrophysical object under observation.

An intriguing advancement of our research could be achieved by first aiming to derive the tidal and rotational Love numbers within the context of rapid rotation. This endeavor is particularly relevant from the perspective of GW astronomy, though it presents a considerable challenge due to its reliance on perturbing the fully rotating metric as a foundational approach. Following this, exploring universal relations for more exotic compact objects, such as spinning mixed scalar boson-fermion stars \cite{Mourelle:2024qgo} or Proca-fermion stars, would be a valuable extension. Additionally, investigating scenarios where a horizon forms within rotating Boson Stars—resulting in hairy Kerr black holes—would further broaden our understanding of universal relations in these complex systems.

We expect that our analysis will become a useful tool for the identification of astrophysical compact objects, primarily with the degeneracy problem in the context of gravitational waveforms of future binary merger events, but also for the search for possible bosonic scalar and vectorial terms for dark matter candidates, and in the further understanding of the strong gravity regime of General Relativity.

\begin{acknowledgements}
 JCM thanks E.Radu for their crucial help with the FIDISOL/CADSOL package and also thanks M.Huidobro and A.G. Martín-Caro for further useful comments.  
Further, the authors acknowledge the Xunta de Galicia (Grant No. INCITE09.296.035PR and Centro singular de investigación de Galicia accreditation 2019-2022), the Spanish Consolider-Ingenio 2010 Programme CPAN (CSD2007-00042), and the European Union ERDF.
AW is supported by the Polish National Science Centre,
grant NCN 2020/39/B/ST2/01553. 
This work is supported by the Center for Research and Development in Mathematics and Applications (CIDMA) through the Portuguese Foundation for Science and Technology Funda\c c\~ao para a Ci\^encia e a Tecnologia), UIDB/04106/2020, UIDP/04106/2020,
https://doi.org/10.54499/UIDB/04106/2020 
and https://doi.org/10.54499/UIDP/04106/2020.
The authors acknowledge support from the projects
http://doi.org/10.54499/PTDC/FISAST/3041/2020,
http://doi.org/10.54499/CERN/FIS-PAR/0024/2021 and https://doi.org/10.54499/2022.04560.PTDC.  This work has further been supported by the European Horizon Europe staff exchange (SE) programme HORIZON-MSCA2021-SE-01 Grant No. NewFunFiCO101086251. JCM thank the Xunta de Galicia (Consellería de Cultura, Educación y Universidad) for funding their predoctoral activity through \emph{Programa de ayudas a la etapa predoctoral} 2021. JCM thanks the IGNITE program of IGFAE for financial support. E.S.C.F. is supported by the FCT grant PRT/BD/153349/2021 under the IDPASC Doctoral Program. Computations have been partially performed at the Argus cluster at the U. Aveiro.
\end{acknowledgements}

\bibliography{biblio}

\vspace*{25cm}

\begin{appendix}

\section{Equations of interest}
\label{appendix}

We show here some equations of interest.
Our BS are obtained by solving the EKG system of equations \cref{kg}, in the following form:
\begin{equation}
\begin{split}
&-e^{2\alpha}\frac{r^2}{2}\sin^2(\theta)\left(-E_{t}^{t}+E_{r}^{r}+E_{\theta}^{\theta}-E_{\psi}^{\psi}\right)=0\\
&e^{2\alpha}\frac{r^2}{2}\sin^2(\theta)\left(E_{t}^{t}+E_{r}^{r}+E_{\theta}^{\theta}-E_{\psi}^{\psi}+\frac{2WE_{\psi}^{t}}{r}\right)=0\\
&e^{2\alpha}\frac{r^2}{2}\sin^2(\theta)\left(-E_{t}^{t}+E_{r}^{r}+E_{\theta}^{\theta}-E_{\psi}^{\psi}-\frac{2WE_{\psi}^{t}}{r}\right)=0\\
&2re^{2\nu +2\alpha-2\beta}E_{\psi}^{t}=0\\
& \frac{e^{2\alpha}r^2\sin^2(\theta)}{\phi} \Phi^*\left(\Box-\frac{d V}{d|\phi|^2} \right) \Phi=0.
\end{split}
\label{EKG-system}
\end{equation}

The Einstein Proca system, well described in \cref{pr}, due to the way we need the field equations for numerical reasons, was obtained using both the usual Einstein field formalism and the Euler-Lagrange method. The metric function equations were obtained as in the above case, within the frame of the General Relativity. To obtain the equations for the  Proca field, rather than use the Einstein field formulation, we have defined an effective lagrangian as follows:
\begin{equation}
 \mathcal{L}_{eff}^P  =\sqrt{-g}\left(-\frac{1}{4}\textit{F}_{\alpha\beta}\bar{\textit{F}}^{\alpha\beta}-\frac{1}{2}\mu^2\textit{A}_{\alpha}\bar{\textit{A}}^{\alpha}+\frac{\mathcal{L}^2}{2}\right) ,
\end{equation}
And the Euler-Lagrange equations are obtained as usual:
\begin{equation}
    \begin{split}
        &ELV=\frac{\partial}{\partial r}\left(\frac{\partial \mathcal{L}_{eff}^P }{\partial V_{,r} }\right)+\frac{\partial}{\partial \theta}\left(\frac{\partial \mathcal{L}_{eff}^P }{\partial V_{,\theta} }\right)-\frac{\partial \mathcal{L}_{eff}^P }{\partial V}=0,\\
        &ELH_i=\frac{\partial}{\partial r}\left(\frac{\partial \mathcal{L}_{eff}^P }{\partial H_{i,r} }\right)+\frac{\partial}{\partial \theta}\left(\frac{\partial \mathcal{L}_{eff}^P }{\partial H_{i,\theta} }\right)-\frac{\partial \mathcal{L}_{eff}^P }{\partial H_i}=0.
    \end{split}
\end{equation}
Where $i=1,2,3.$
The combinations of equations that allow us to have the Einstein-Proca system described properly for entry in the solver, is the following:

\begin{equation}
\begin{split}
&e^{2\alpha}\frac{r^2}{2}\sin^2(\theta)\left(E_{t}^{t}-E_{r}^{r}-E_{\theta}^{\theta}+E_{\psi}^{\psi}\right)=0,\\
&e^{2\alpha}\frac{r^2}{2}\sin^2(\theta)\left(E_{t}^{t}+E_{r}^{r}+E_{\theta}^{\theta}-E_{\psi}^{\psi}+\frac{2WE_{\psi}^{t}}{r}\right)=0,\\
&e^{2\alpha}\frac{r^2}{2}\sin^2(\theta)\left(-E_{t}^{t}+E_{r}^{r}+E_{\theta}^{\theta}-E_{\psi}^{\psi}-\frac{2WE_{\psi}^{t}}{r}\right)=0,\\
&2re^{2\nu +2\alpha-2\beta}E_{\psi}^{t}=0,\\
& e^{\beta-\nu}r\sin\theta\left(W\sin\theta  ELV-ELH_3\right)=0,\\
&e^{-\beta+\nu}\sin\theta\left[ELV+e^{-2\nu+2\beta}\sin\theta W\left(ELH_3\right.\right.\\&
\left.\left.-W\sin\theta ELV\right)\right]=0,\\
&-e^{\nu+2\alpha-\beta}r^2\sin\theta\left(ELH_2+e^{\nu+\beta}H_2\sin\theta E^r_r\right)=0,\\
&-e^{\nu+2\alpha-\beta}r^2\sin\theta\left(ELH_1+e^{\nu+\beta}H_1\sin\theta E^{\theta}_{\theta}\right)=0.
\end{split}
\label{EP-system}
\end{equation}

We also add the explicit formulation for the Noether current components for the vectorial cases, as the moment of Inertia is obtained using the ratio between them \cref{diferentialfrequencyP}

\begin{equation}
\begin{split}
    &j^t_p=\frac{e^{-2( \nu+ \alpha)}}{r^2} \left[\sin \theta  W \left(H_1 H_{3,r}+\frac{H_2 H_{3,\theta}}{r}\right)\right.\\&
    \left.+\left(H_1^2+H_2^2\right) \left(w-\frac{m W}{r}\right)+r H_1 V_{,r}\right.\\&
    \left.+H_2 V_{,\theta}+\frac{\cos \theta  H_2 H_3 W}{r}\right]+\frac{e^{-2 \nu-2 \beta}}{r^2} \left[m \csc \theta  H_3 V+w H_3^2\right];
\end{split}
\end{equation}

\begin{equation}
    \begin{split}
    &j^{\psi}_p=\frac{W e^{-2 (\nu+\alpha)}}{r^4} \left[r H_1 \left(\sin \theta  H_{3,r} W+r V_{,r}\right)-m H_1^2 W\right.\\&
    \left.+H_2 \left\{W \left(-m H_2+\sin \theta  H_{3,\theta}+\cos \theta  H_3\right)+r V_{,\theta}\right\}\right]\\
    &-\frac{e^{-2 (\nu+\alpha+\beta)} }{r^3}\left\{w e^{2 \beta} W \left(H_1^2+H_2^2\right)\right.\\&
    \left.-r \csc \theta  e^{2 \alpha} V (w H_3+m \csc \theta  V)\right\}\\&
    +\frac{\csc \theta  e^{-2 (\alpha+\beta)}}{r^4} \left(r H_1 H_{3,r}+H_2 H_{3,\theta}\right)+\\
    &\frac{e^{-2 (\alpha+\beta)} }{r^4}\left(m \csc ^2\theta  \left(H_1^2+H_2^2\right)-\cot \theta  \csc \theta  H_2 H_3\right).
    \end{split}
\end{equation}

\vspace{0.5cm}
{\bf{Numerical parameters used:}}
In this paper, we have selected a set of physically well-motivated potentials, fitting various astrophysical scenarios, like dark matter haloes, \cite{Mielke:2019rvl,Mielke:2020mve}, BH and NS-like objects \cite{Choi:2019mva,Guerra:2019srj,Delgado:2020udb,Vaglio:2022flq,Grandclement:2014msa}All of them considered in the literature and arising different qualitative BSs' features.

\begin{table}[h!]
	\centering
		\begin{tabular}{|c|c|}
			\hline
		 Name & $V\left(\phi\right)$ \\ \hline
			Mini-BS, BS$_{\rm Mass}$& $V_{\rm Mass}=\mu^2\phi^2$ \\ \hline
		BS$_{\rm Quartic}$&$V_{\rm Quartic}=\mu^2\phi^2+|\lambda|/2\phi^4$    \\  \hline
				BS$_{\rm Halo}$& $V_{\rm Halo}=\mu^2\phi^2-|\alpha|\phi^4$  \\ \hline
			BS$_{\rm HKG}$ & $V_{\rm HKG}=\mu^2\phi^2-\alpha\phi^4+\beta\phi^6$  \\ \hline
			BS$_{\rm Sol}$& $V_{\rm Sol}=\mu^2\phi^2(1-(\phi^2/\phi_0^2))^2$  \\ \hline
          BS$_{\rm Sant}$& $V_{\rm Sant}=\mu^2\phi^2(1-(\phi^4/\phi_0^2))^2$  \\ \hline
			BS$_{\rm Log}$&$V_{\rm Log}=f^2\mu^2 \ln\left(\phi^2/f^2+1\right)$\\ \hline
			BS$_{\rm Liouville}$& $V_{\rm Liouville}=f^2\mu^2 \left(\exp\{\phi^2/f^2\}-1\right)$  \\ \hline
				BS$_{\rm Axion}$& $V_{\rm Axion}=\frac{2\mu^2f^2}{B}\left(1-\sqrt{1-4B\sin^2(\phi/2f)}\right)$  \\
			\hline
		\end{tabular}
		\caption{\small BS potentials analyzed in the current work have been previously considered in the case
of spherical, non-rotating BSs, and some cases, further
generalized to rotating solutions   \cite{Siemonsen:2020hcg}. 
Ranging from the so-called \textit{Mini-boson star} potential, through the inclusion of higher order self-interaction terms, e.g. $|\Phi|^4$ and $|\Phi|^6$ \cite{Schunck:2003kk,Colpi:1986ye,Grandclement:2014msa}.Also potentials based on the logarithm, exponential, sine functions and the axion potential\cite{Delgado:2020udb,Choi:2019mva,Guerra:2019srj}. }
		\label{Table.Potentials}
\end{table}

Here we give the different numerical sets of values for the parameters that we have used for our simulations. The numerical values are given in rescaled units.
We start showing the BS models:

 \begin{equation}
V_{\rm Quartic}=\phi^2+\frac{\lambda}{2}\phi^4
\begin{cases}
      \lambda= 1\\
      \lambda=10\\
      \lambda=40\\
      \lambda=50\\
      \lambda=60\\
      \lambda=70\\
      \lambda=80
\end{cases}
 \end{equation}

\begin{equation}
V_{\rm Halo}=\phi^2-\alpha\phi^4
\begin{cases}
      \alpha= 1,\\
      \alpha=12.\\
\end{cases}
 \end{equation}

\begin{equation}
V_{\rm HKG}=\phi^2-\alpha\phi^4+\beta\phi^6
\begin{cases}
      \alpha=80, &\beta=0.01\\
      \alpha=2, &\beta=1.8\\
\end{cases}
 \end{equation}

\begin{equation}
V_{\rm Sant}=\phi^2\left(1-\left(\frac{\phi^4}{\phi_0^2}\right)\right)^2
\begin{cases}
      \phi_0=1.5, \\
      \phi_0=0.7,\\
      \phi_0=0.3,\\
       \phi_0=0.1,\\
        \phi_0=0.05.
\end{cases}
 \end{equation}

\begin{equation}
V_{\rm Sol}=\phi^2\left(1-\left(\frac{\phi^2}{\phi_0^2}\right)\right)^2
\begin{cases}
      \phi_0=1.5, 
\end{cases}
 \end{equation}

\begin{equation}
\begin{split}
V_{\rm Axion}&=\frac{2f^2}{B}\left(1-\sqrt{1-4B\sin^2(\frac{\phi}{2f})}\right)\\&
\begin{cases}
      f=0.1, &B=0.22.\\
      f=0.05, &B=0.22.
\end{cases}
\end{split}
 \end{equation}

\begin{equation}
V_{\rm Log}=f^2 \ln\left(\phi^2/f^2+1\right)
\begin{cases}
      f=0.7, \\
      f=0.5.\\
\end{cases}
 \end{equation}

\begin{equation}
V_{\rm Liouville}=f^2 \left(e^\frac{\phi^2}{f^2}-1\right)
\begin{cases}
      f=0.8. \\
\end{cases}
 \end{equation}

The PHS models:

\begin{table}[h!]
	\centering
		\begin{tabular}{|c|c|}
			\hline
		 Models & Parameters \\ \hline
			Model $1$& $\alpha=0.35$, $\Lambda=1$ \\ \hline
		Model $2$&$\alpha=0.50$, $\Lambda=1$    \\  \hline
				Model $3$& $\alpha=5.0$, $\Lambda=1$   \\ \hline
			Model $4$ & $\alpha=10.0$, $\Lambda=1$  \\ \hline
			Model $5$& $\alpha=1$, $\Lambda=0.005$  \\ \hline
          Model $6$& $\alpha=1$, $\Lambda=0.5$ \\ \hline
			Model $7$& $\alpha=1$, $\Lambda=1.0$  \\ \hline
				Model $8$& $\alpha=1$, $\Lambda=5.0$   \\ \hline
				Model $9$& $\alpha=1$, $\Lambda=0.1$   \\
			\hline
		\end{tabular}
		\caption{Different Proca-Higgs stars models obtained varying the potential parameters. Recall that $v$ is related to $\alpha$ through the equation $\alpha^2\equiv 4\pi G v^2$, as previously shown in \Cref{numerical}.  }
		\label{Table.Potentials}
\end{table}
   
\newpage
   
\vspace{0.5cm}

\section{Fitting coefficients}
\label{appendixB}

\begin{widetext}

\begin{table}[h!]
\centering
\begin{tabular}{|ll|l|l|}
\hline
\multicolumn{2}{|l|}{Coeffs}     & $ A_0=0.9800$ & $B=0.9933$ \\ \hline
\multicolumn{1}{|l|}{$A_{1}^{0}=9.4471$} & $A_{2}^{0}=-49.4872$ & $A_{4}^{0}=44.0732$ & $A_{6}^{0}=-6.4055$ \\ \hline
\multicolumn{1}{|l|}{$A_{1}^{1}=-27.6102$} & $A_{2}^{1}=134.0676$ & $A_{4}^{1}=-115.6019$ & $A_{6}^{1}=16.8475$ \\ \hline
\multicolumn{1}{|l|}{$A_{1}^{2}=32.2443$} & $A_{2}^{2}=-142.8045$ & $A_{4}^{2}=118.7492$ & $A_{6}^{2}=-17.2574$ \\ \hline
\multicolumn{1}{|l|}{$A_{1}^{3}=-18.4678$} &$A_{2}^{3}=74.5797$ & $A_{4}^{3}=-59.6621$  &$A_{6}^{3}=8.6004$  \\ \hline
\multicolumn{1}{|l|}{$A_{1}^{4}=5.1696$} &$A_{2}^{4}=-19.0730$  & $A_{4}^{4}=14.6520$ & $A_{6}^{4}=-2.0843$ \\ \hline
\multicolumn{1}{|l|}{$A_{1}^{5}=-0.5637$} &$A_{2}^{5}=1.9101$  & $A_{4}^{5}=-1.4040$   & $A_{6}^{5}=0.1968$       \\ \hline
\end{tabular}
\caption{Numerical values of the coefficients that fit the
universal $s_3-\chi-\alpha$ surface for $n=1$ PS and PHS.
}
\end{table}

\begin{table}[h!]
\centering
\begin{tabular}{|lll|l|l|}
\hline
\multicolumn{3}{|l|}{Coeffs}                                        & $ A_0=-29.5332$ &$B=0.9216$  \\ \hline
\multicolumn{1}{|l|}{$A_{0}^{0}=56.0630$} & \multicolumn{1}{l|}{$A_{1}^{0}=-60.0463$} & $A_{2}^{0}=51.3745$ & $A_{3}^{0}=-19.3585$ &$A_{4}^{0}=2.7020$  \\ \hline
\multicolumn{1}{|l|}{$A_{0}^{1}=-29.6509$} & \multicolumn{1}{l|}{$A_{1}^{1}=47.8969$} & $A_{2}^{1}=-28.8138$ & $A_{3}^{1}=7.5695$ & $A_{4}^{1}=-0.7327$ \\ \hline
\multicolumn{1}{|l|}{$A_{0}^{2}=-2199.5518$} & \multicolumn{1}{l|}{$A_{1}^{2}=5468.2548$} & $A_{2}^{2}=-5002.2628$ & $A_{3}^{2}=2002.0281$ &$A_{4}^{2}=-297.6725$  \\ \hline
\multicolumn{1}{|l|}{$A_{0}^{4}=65416.3487$} & \multicolumn{1}{l|}{$A_{1}^{4}=-170451.166$}  & $A_{2}^{4}=165032.745$ &  $A_{3}^{4}=-70539.648$  &$A_{4}^{4}=11379.1935$  \\ \hline
\multicolumn{1}{|l|}{$A_{0}^{6}=56.0630$} & \multicolumn{1}{l|}{$A_{1}^{6}=660143.052$}  &$A_{2}^{6}=-1.1846913e6$   & $A_{3}^{6}=7.1420735e5$   &$A_{4}^{6}=-1.4988229e5$  \\ \hline
\multicolumn{1}{|l|}{$A_{0}^{8}=-1.60634751e7$}  & \multicolumn{1}{l|}{$A_{1}^{8}=2.70723625e7$}   & $A_{2}^{8}=-1.48794001e7$  &$A_{3}^{8}=2.37628246e6$    &$A_{4}^{8}=2.40259493e5$  \\ \hline
\end{tabular}
\caption{Numerical values of the coefficients that fit the
universal $\textit{C}-\chi-\alpha$ surface for $n=1$ PS and PHS.
We use the notation $en=10^n$ being $n\in\mathcal{N}$.}
\end{table}

\begin{table}[h!]
\centering
\begin{tabular}{|llll|l|l|}
\hline
\multicolumn{4}{|l|}{Coeffs}                                                                      &$ A_0=-644.206$ & $B=0.4165$ \\ \hline
\multicolumn{1}{|l|}{$A_{1}^{0}=-5.5070432e3$} & \multicolumn{1}{l|}{$A_{2}^{0}=-1.60429048e4$} & \multicolumn{1}{l|}{$A_{3}^{0}=-1.70219602e4$} & $A_{4}^{0}=-2.5619378e3$ & $A_{5}^{0}=-1.122663e3$ & $A_{6}^{0}=-8.561946e3$ \\ \hline
\multicolumn{1}{|l|}{$A_{1}^{1}=-1.815226e3$} & \multicolumn{1}{l|}{$A_{2}^{1}=-1.68859476e4$} & \multicolumn{1}{l|}{$A_{3}^{1}=-5.42111748e4$} & $A_{4}^{1}=-6.77827317e4$ & $A_{5}^{1}=-1.98600867e4$ & $A_{6}^{1}=1.57278199e4$  \\ \hline
\multicolumn{1}{|l|}{$\sim$} & \multicolumn{1}{l|}{$A_{2}^{2}=-9.99916e2$} & \multicolumn{1}{l|}{$A_{3}^{2}=-1.1487998e4$} & $A_{4}^{2}=-4.11673438e4$ & $A_{5}^{2}=-5.61465386e4$ & $A_{6}^{2}=-2.91234017e4$   \\ \hline
\multicolumn{1}{|l|}{$\sim$} & \multicolumn{1}{l|}{$\sim$} & \multicolumn{1}{l|}{$A_{3}^{3}=3.171403e2$} & $A_{4}^{3}=4.071305e2$ & $A_{5}^{3}=-3.8345522e3$ & $A_{6}^{3}=-2.623955e3$ \\ \hline
\multicolumn{1}{|l|}{$\sim$} & \multicolumn{1}{l|}{$\sim$} & \multicolumn{1}{l|}{$\sim$} & $A_{4}^{4}=1.913004e2$ & $A_{5}^{4}=1.0913813e3$ & $A_{6}^{4}=10.0917$  \\ \hline
\multicolumn{1}{|l|}{$\sim$} & \multicolumn{1}{l|}{$\sim$} & \multicolumn{1}{l|}{$\sim$} & $\sim$ & $A_{5}^{5}=-23.8479$ & $A_{6}^{5}=205.6910$  \\ \hline
\multicolumn{1}{|l|}{$\sim$} & \multicolumn{1}{l|}{$\sim$} & \multicolumn{1}{l|}{$\sim$} & $\sim$ & $\sim$ & $A_{6}^{6}=-18.8144$ \\ \hline
\end{tabular}
\caption{Numerical values of the coefficients that fit the
universal $\beta-\chi-\alpha$ surface for $n=1$ PS and PHS.We use the notation $en=10^n$ being $n\in\mathcal{N}$.
}
\end{table}

\end{widetext}
\end{appendix}

\end{document}